\documentclass[ reprint,amsmath,amssymb,aps]{revtex4-2}

\usepackage[tight,footnotesize]{subfigure}
\usepackage[justification=justified]{caption}
\usepackage[font=footnotesize]{subfig}
\usepackage{graphicx}
\usepackage{amsmath,amssymb,amsfonts}
\usepackage{color,soul}
\usepackage{dsfont}
\usepackage{xcolor}
\usepackage{soul}

\usepackage{epsfig}
\usepackage{psfrag}
\usepackage{epstopdf}
\usepackage{pstool}

\usepackage[newcommands]{ragged2e}
\usepackage{graphicx,caption}

\newcommand{\dya}[1]{\bar{\bar{#1}}}

\begin{document}

\preprint{APS/123-QED}

\title{Transmittable Nonreciprocal Cloaking}
\author{Mojtaba Dehmollaian}
\author{Guillaume Lavigne}%
\affiliation{Poly-Grames Research Center, Polytechnique Montréal, 2500 Chem. de Polytechnique, H3T 1J4, Montréal, Québec, Canada}
\author{Christophe Caloz}
\affiliation{ESAT-WAVECORE-META Research Center, KU Leuven, Kasteelpark Arenberg 10, 3001, Leuven, Belgium}
\date{\today}

\begin{abstract}
Cloaking is typically reciprocal. We introduce here the concept of \emph{transmittable nonreciprocal cloaking} whereby the cloaking system operates as a standard omnidirectional cloak for external illumination, but can transmit light from its center outwards at will. We demonstrate a specific implementation of such cloaking that consists in a set of concentric bianisotropic metasurfaces whose innermost element is nonreciprocal and designed to simultaneously block inward waves and pass -- either omnidirectionaly or directionally -- outward waves. Such cloaking represents a fundamental diversification of conventional cloaking and may find applications in areas such as stealth, blockage avoidance, illusion and cooling. 
\end{abstract}

\maketitle


\section{\label{sec:introduction}Introduction}

Cloaking is a powerful concept in electromagnetics that has emerged in 2006 as an outgrowth of metamaterials~\cite{leonhardt2006optical,pendry2006controlling} and that has known vibrant and unabated development since then~\cite{fleury2015review}. A cloak is a metamaterial shell structure with medium properties that are designed so as to curve the trajectory of incident light around its core, whose contents is hence made invisible to external observers. Different types of cloaking techniques have been reported, including coordinate-transformation deviation~\cite{pendry2006controlling, schurig2006metamaterial}, scattering cancellation~\cite{alu2005plasmonic, silveirinha2007parallel}, transmission-line matching~\cite{alitalo2009electromagnetic}, gain compensation~\cite{Selvanayagam2013active} and metasurface multiple scattering~\cite{dehmollaian2021concentric} or waveguiding~\cite{lee2022metasurface}. 

The quasi-totality of the cloaking structures reported to date are \emph{reciprocal}: they fully satisfy the Lorentz reciprocity theorem~\cite{Lorentz_1896}. Exceptions are the one-way cloaks presented in~\cite{he2011one,zhu2013one} and the unidirectional loss-and-gain balanced cloak presented in~\cite{sounas2015unidirectional}. These devices perform cloaking for light incident from a given direction, but reflect light incident from the opposite direction. They represent therefore \emph{nonreciprocal} cloaks. 

We present here a completely different type of nonreciprocal cloak. This device exhibits the property of \emph{transmittable nonreciprocity}, operating as a standard -- and hence also omnidirectional ({contrarily to}~\cite{he2011one,zhu2013one,sounas2015unidirectional}) -- cloak for external illumination, and as a transmission medium, activable at will and allowing beam forming, for internal (core) illumination. It is implemented in the form of a set of concentric bianisotropic metasurfaces~\cite{lavigne2018susceptibility,achouri2021electromagnetic} with the innermost element being a nonreciprocal metasurface, which may be realized in magnetless transistor technology~\cite{lavigne2022metasurface}.


\section{\label{sec:principle}Operation Principle}

\begin{figure}[h!]
	\includegraphics[width=1\columnwidth]{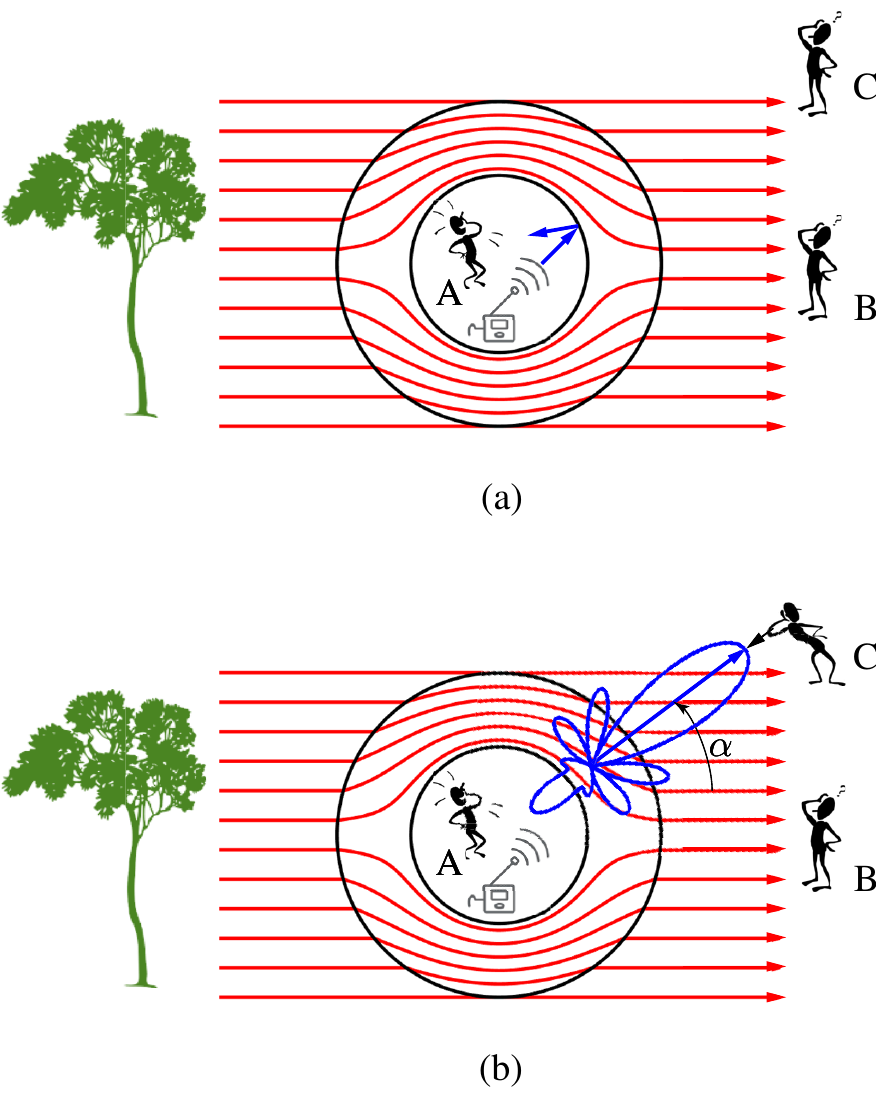}
	\caption{\label{fig:concept}Comparative description of transmittable nonreciprocal cloaking. (a)~Conventional (reciprocal) cloaking, with invisibility for external illumination and reflection for internal illumination. (b)~Proposed nonreciprocal cloaking, with conventional (omnidirectional) invisibility for external illumination and transmission -- possibly directive -- for internal illumination.}
\end{figure}

Figure~\ref{fig:concept} provides a comparative description of the proposed transmittable nonreciprocal cloaking concept, with the usual cloaking shell structure and its core whose contents is made invisible, via light deviation, to external observers for external illumination.

Figure~\ref{fig:concept}(a) depicts reciprocal cloaking. Waves (red arrows) impinging on the structure from an external source are bent by the cloaking shell around the core, A, which is hence made invisible to external observers, B and C. While the figure represents an incident \emph{plane wave}, with trivial angular spectrum $\delta(\vec{k}-\vec{k}_\text{i})$, where $\vec{k}_\text{i}$ is the (fixed) incident wavevector, the cloaking effect occurs for \emph{any type of wave} (e.g., circular wave emitted by a close point source) and for any \emph{incidence angle} (circular symmetry: $\theta_\text{i}\in\left[-\pi/2,\pi/2\right]$, or omnidirectionality). None of the rays forming the incident wave, whatever its nature, can penetrate into the core region, given its point-singularity origin. Therefore, in the absence of an external force and nonlinearity~\cite{Caloz_PRAp_10_2018}, light emitted from the core region would also not find any transmission channel through the cloaking shell, and would hence be reflected from it, as illustrated in the figure (blue arrows). The system is thus fully reciprocal.

Figure~\ref{fig:concept}(b) presents the proposed concept of \emph{transmittable nonreciprocal cloaking}. The operation of the system is identical to that of the conventional cloak [Fig.~\ref{fig:concept}(a)] for \emph{external illumination}, i.e., the system cloaks its contents for any incident wave and incidence angle, making A invisible to observers B and C. However, instead of always reflecting light for \emph{internal illumination}, the system can directionally \emph{transmit} light through the shell outwards to an intended external observer, C, who would then see either the background environment, as B, if A is silent, or a superposition of the background environment \emph{and} a wave coming from the core of the system if A emits.

\section{\label{sec:implementation}Concentric Metasurface Implementation}

Constructing a cloak, even reciprocal, is generally a challenging task. The most powerful cloaking technique -- coordinate transformation -- requires a complex voluminal inhomogeneous and anisotropic medium as well as unattainable infinite parameter values at the innermost boundary of the shell~\cite{pendry2006controlling}, while other cloaking techniques involve well-documented other difficulties along with specific limitations. On the other hand, both reciprocal~\cite{lavigne2018susceptibility} \emph{and} nonreciprocal metasurfaces~\cite{travati2018nonreciprocal,lavigne2022metasurface} have been recently demonstrated as practically viable electromagnetic devices. Therefore, we select here a metasurface-based approach for the implementation of the transmittable nonreciprocal cloak in Fig.~\ref{fig:concept}(b).

\begin{figure}[h!]
	\includegraphics[width=1\columnwidth]{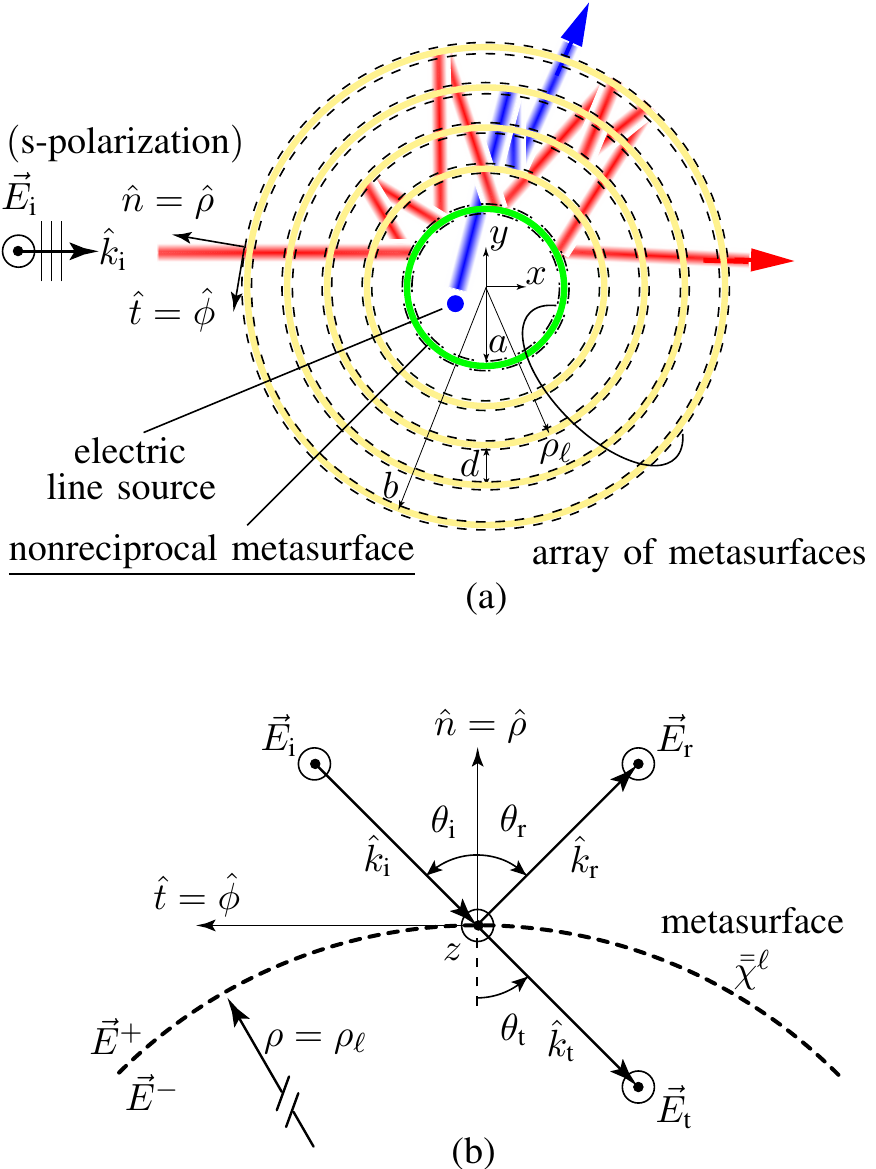}
	\caption{\label{fig:structure}Metasurface implementation of the transmittable nonreciprocal cloak in Fig.~\ref{fig:concept}(b). (a)~Concentric metasurface structure. (b)~Curved metasurface sheet modeling any of the metasurfaces in (a).}
\end{figure}

Figure~\ref{fig:structure} describes the selected metasurface implementation structure. This structure, whose reciprocal version was initially suggested in~\cite{dehmollaian2021concentric}, consists in a set of concentric uniform circular bianisotropic (gainless and lossless) metasurfaces with the innermost element replaced by a nonreciprocal metasurface, as shown in Fig.~\ref{fig:structure}(a). The overall assembly forms a multiple-scattering system akin to a circular (cylindrical or spherical) {metasurface-enhanced} multilayer Fabry-Perot resonator that is optimized for minimal scattering (maximal cloaking) and nonreciprocity (outward transmission), leveraging the great parametric diversity of the system, which includes arbitrary magnitude and phase of the reflection (generally asymmetric) and transmission parameters at each of the metasurfaces as well as arbitrary interspacing between the metasurfaces. The detailed design procedure will be presented in Sec.~\ref{subsec:overall_str}.

{Given its circular cavity, bianisotropic interface and radially nonuniform features, this Fabry-Perot structure seems too complex to admit a precise explanation of the cloaking operation in terms of simple physics. However, this operation may be qualitatively understood as optimal wave routing along the porous circular waveguides formed by the metasurfaces, as suggested in Fig.~\ref{fig:structure}(a) and as will be later illustrated in full-wave simulations~\footnote{As all cloaks, this cloaking system is subjected to a fundamental trade-off between cloaking efficiency (minimal scattering) and its operation bandwidth, due to the physical impossibility (or impracticality~\cite{jackson1999classical}) for the deflected part of the energy near the center to propagate superluminally for proper phase synchronization with the undeflected part of the energy in the periphery~\cite{miller2006perfect}.}. Fortunately, the nonreciprocal operation does not add major complexity to the operation of the overall system. Indeed, as will be seen in Sec.~\ref{sec:design}, given the inner boundary location and inward penetrability of the nonreciprocal metasurface, the cloaking design will be independent from the transmission design, while the transmission design will only unrestrictively depend on the cloaking design, with the related nonreciprocal metasurface design following standard transistor-loaded~\cite{travati2018nonreciprocal,Lavigne_NJP_07_2021,taravati2021full,lavigne2022metasurface} or time-modulated~\cite{shi2017optical,taravati2020full} nonreciprocal metasurface technologies.}


\section{\label{sec:ms_model}Metasurface Modeling}
%
The metasurfaces constituting the transmittable nonreciprocal cloaking structure in Fig.~\ref{fig:structure}(a) are generically represented, with relevant parameters, in Fig.~\ref{fig:structure}(b). We model here these metasurfaces via the Generalized Sheet Transition Conditions (GSTCs)~\cite{idemen1987boundary,kuster2003averaged,achouri2015general}, which are a generalization of the classical boundary conditions including bianisotropic surface polarization current densities~\cite{achouri2021electromagnetic}. In the modeling, it is assumed that the radius of curvatures of the metasurfaces are large compared to the wavelength so that the incident waves locally see homogeneous flat sheets and hence negligible diffraction effects.

Assuming an s-polarization scenario (two-dimensional problem), zero normal surface currents (for simplicity) and the harmonic time convention $e^{\text{j} \omega t}$, the GSTCs read (see Supp. Mat.~\ref{appendix:model})
\begin{subequations}\label{eq:GSTC_susceptibility}
	\begin{equation}
	E_z^{+}-E_z^{-} = \text{j}k_0\left(\chi_{\text{me}}^{\phi z}E_{z,\text{av}}+\eta_0\chi_{\text{mm}}^{\phi \phi}H_{\phi,\text{av}}\right)
	\end{equation}        
	and                                 
	\begin{equation}
	H_\phi^{+}-H_\phi^{-} =\frac{\text{j}k_0}{\eta_0}\left(\chi_{\text{ee}}^{z z}E_{z,\text{av}}+\eta_0\chi_{\text{em}}^{z\phi }H_{\phi,\text{av}}\right),
	\end{equation}
\end{subequations}
where superscripts $\pm$ refer to the fields at $\rho=\rho_\ell^\pm$, just above and below the $\ell^\text{th}$ sheet, $k_0$ and $\eta_0$ are the free-space wavenumber and wave impedance, respectively,  
${\chi}_{\text{ee}}^{zz}$, $\chi_{\text{em}}^{z\phi}$, $\chi_\text{me}^{\phi z}$ and $\chi_\text{mm}^{\phi \phi}$ represent electric-to-electric, magnetic-to-electric, electric-to-magnetic and magnetic-to-magnetic surface susceptibilities, respectively, and $E_{z,\text{av}}=(E_z^{+} + E_z^{-})/2$ and $H_{\phi,\text{av}}=(H_\phi^{+} + H_\phi^{-})/2$ denote the average electric and magnetic fields at the metasurface sheet, respectively. The susceptibilities ${\chi}_{\text{ee}}^{zz}$, $\chi_{\text{em}}^{z\phi}$, $\chi_\text{me}^{\phi z}$ and $\chi_\text{mm}^{\phi \phi}$ correspond to the assumed s-polarization regime, with the electric field along $\hat{z}$ and the tangential magnetic filed along $\hat{\phi}$; in the p-polarized case, the relevant susceptibilities, corresponding to the tangential electric field along $\hat{\phi}$ and the magnetic field along $\hat{z}$, would be ${\chi}_{\text{ee}}^{\phi\phi}$, $\chi_{\text{em}}^{\phi z}$, $\chi_\text{me}^{z\phi}$, and $\chi_\text{mm}^{zz}$.  

The $\ell^\mathrm{th}$ metasurface will be denoted by the same superscript, as indicated in Fig.~\ref{fig:structure}, and, according to Eqs.~\eqref{eq:GSTC_susceptibility}, the corresponding (s-polarization) susceptibility will be written in the compact tensorial form
\begin{equation}\label{eq:susc_tens}
\dya{\chi}^\ell = {\chi_{\text{ee}}^{zz,\ell}}\hat{z}\hat{z}+ {\chi_{\text{em}}^{z\phi,\ell}} \hat{z}\hat{\phi} + {\chi_{\text{me}}^{\phi z,\ell}}\hat{\phi}\hat{z} + 
{\chi_{\text{mm}}^{\phi\phi,\ell}} \hat{\phi}\hat{\phi},
\end{equation}
where the four susceptibilities are constant, i.e., not functions of $\phi$, according to the uniformity (or circular symmetry) assumption that ensures cloaking omnidirectionality.

\section{\label{sec:design}Cloak Design}

\subsection{\label{subsec:overall_str}{Overall} Procedure}

\begin{figure}[h!]
	\includegraphics[width=1\columnwidth]{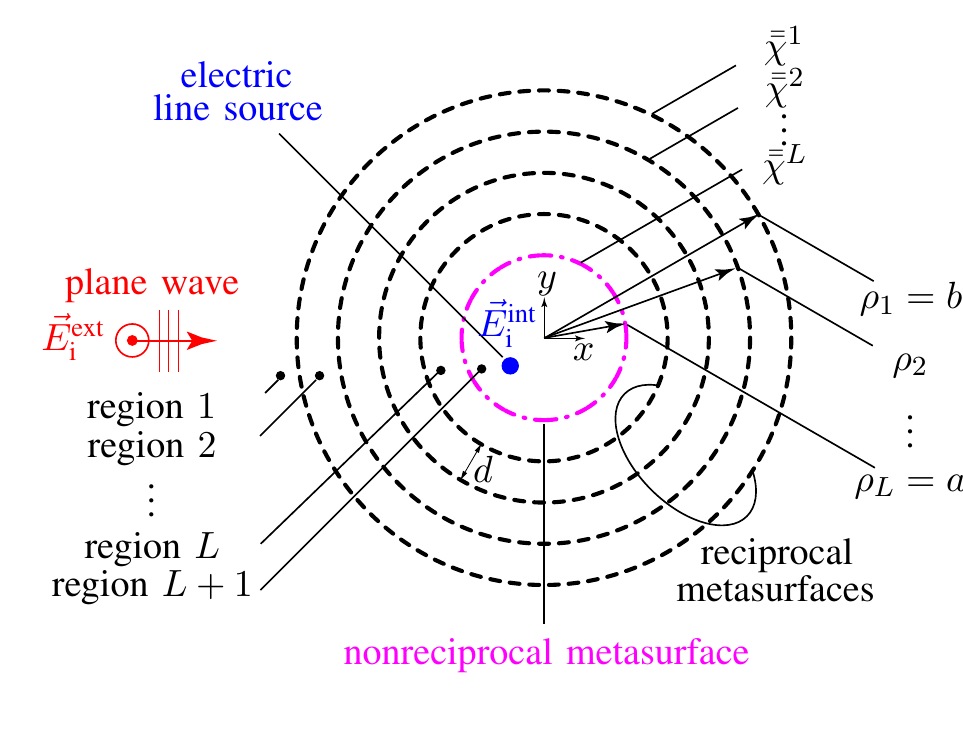}
	\caption{\label{fig:design}Parametric setup for the design of the metasurface-based transmittable nonreciprocal cloak in Fig.~\ref{fig:structure}(a).}
\end{figure}

%
{We shall design the transmittable nonreciprocal cloak in Fig.~\ref{fig:structure}(a) by successively optimizing the structure for cloaking in the external plane-wave illumination regime and for transmission in the internal point/line-source illumination regime, based on the parametric setup shown in Fig.~\ref{fig:design}. This will be accomplished by using the electromagnetic analysis tool to be established in Sec.~\ref{subsec:analysis}, which, incorporating the bianisotropic-susceptibility GSTC metasurface model presented in Sec~\ref{sec:ms_model}, provides the exact electromagnetic fields everywhere in the system, while the optimization can be performed with any standard optimization tool.}

{The twofold cloaking-transmission optimization will result in the determination of the $4L$ susceptibility parameters in Eq.~\eqref{eq:susc_tens}, for a given core radius, $a$, and cloak radius, $b$, assuming, for simplicity, uniform metasurface interspacing, $d$. The cloaking optimization will fully determine the susceptibility parameters of the metasurfaces $1,\ldots,L-1$, and part of the susceptibility parameters of the metasurface $L$, while the nonreciprocity optimization will determine the remaining susceptibility parameters of the metasurface $L$.}


{The cloaking and transmission designs, as announced in Sec.~\ref{sec:implementation}, are essentially independent from each other, as far as the metasurfaces are concerned. Such independence will be ensured by specifically making the $L^\mathrm{th}$ metasurface impenetrable to external illumination, i.e., by imposing $\left|S_{11}^L\right|=1$, while subjecting the corresponding phase, $\angle S_{11}^L$ to cloaking optimization. In the internal-illumination regime, the cloaking-optimized globally transmissive nature of the metasurfaces $1,\ldots,L-1$ will automatically provide an exit channel to the transmitted wave, while its radiation features may be independently controlled using antenna design principles. The case of simultaneous external and internal illumination will be discussed in Sec.~\ref{subsec:external_internal_illuminations}.}

\subsection{\label{subsec:analysis}Electromagnetic Analysis}

We analyze the system in Fig.~\ref{fig:design} by successively expanding the metasurface-tangential fields in the different regions in cylindrical Bessel functions~\cite{harrington2001time}, applying the GSTCs~\eqref{eq:GSTC_susceptibility} at each metasurface interface between these regions~\cite{achouri2021electromagnetic} and resolving the resulting matrix system to obtain the expansion field coefficients.

%
The tangential electric and magnetic fields in the $\ell^\mathrm{th}$ region can be expressed as 
\begin{subequations}\label{eq:fields_expansion}
	\begin{equation}\label{eq:fields_expansion_E}
	E_z^\ell = \sum_{n=-N}^{n=N}\text{j}^{-n}\left[b_n^\ell J_n(k_\ell \rho)+a_n^\ell H_n^{(2)}(k_\ell \rho)\right]\text{e}^{\text{j}n\phi}
	\end{equation}
	and, from the Maxwell-Amp\`{e}re's equation,
	\begin{equation}
	H_\phi^\ell=\frac{1}{\text{j}k_\ell\eta_\ell}\frac{\partial E_z^\ell}{\partial \rho},
	\end{equation}
\end{subequations}
where $J_n(\cdot)$ is the cylindrical Bessel function of the first kind, accounting for multiple scattering within the annular and core regions, $H_n^{(2)}(\cdot)$ is the cylindrical Hankel function of the second kind, accounting for radiation across the interfaces, $k_\ell$ and $\eta_\ell$ are the wavenumber and wave impedance of region $\ell$, and $a_n^\ell$ and $b_n^\ell$ are the corresponding unknown expansion coefficients.  

%
From this point, we enforce the GSTCs~\eqref{eq:GSTC_susceptibility} with the fields~\eqref{eq:fields_expansion} at each metasurface boundary ($\ell=1,2,\dots,L$) and match term-by-term ($n=-N,\dots,N$) the modal contributions of the resulting equations for the aforementioned external and internal illuminations; this leads to a linear matrix system whose solutions are the field expansion coefficients $a^\ell_n$ and $b^\ell_n$ (see Supp. Mat.~\ref{appendix:analysis}).

\subsection{\label{subsec:cloaking}Cloaking-Regime Scattering Minimization}

We make the following assumptions: i)~$\vec{E}^\text{int}_\text{i}=0$ while $\vec{E}^\text{ext}_\text{i}\neq 0$, where $\vec{E}^\text{int}_\text{i}$ and $\vec{E}^\text{ext}_\text{i}$ are the external and internal incident fields, respectively (Fig.~\ref{fig:design}); ii)~all the metasurfaces are reciprocal, except the innermost one ($\ell=L$); iii)~all the metasurfaces are lossless and gainless, except the innermost one (nonreciprocity implies some form of gain~\cite{Caloz_PRAp_10_2018}); iv)~the innermost region (region $L+1$) operates as a perfect electric conductor under external illumination, specifically its permittivity is set to a very large negative imaginary number, to ensure impenetrability of the cloak's core. The reciprocity condition implies that $\chi_\text{em}^{z\phi,\ell}=-\chi_\text{me}^{\phi z,\ell}$, while the gain-less and lossless condition implies that $\chi_\text{ee}^{zz,\ell}$ and $\chi_\text{mm}^{\phi\phi,\ell}$ are purely real and $\chi_\text{em}^{z\phi,\ell}=-\chi_\text{me}^{\phi z,\ell}$ are purely imaginary~\cite{achouri2021electromagnetic}. 

We shall quantify the scattering of the structure under external illumination in terms of the scattering echo width~\cite{ishimaru2017electromagnetic}, namely 
\begin{equation}\label{eq:SEW}
\delta(\phi)=\lim_{\rho\to\infty}2\pi\rho\left|\frac{E^\text{scat}}{E^\text{inc}}\right|^2=\frac{4}{k_0}\left|\sum_{n=-N}^{n=N}a_n^1\text{e}^{\text{j}n\phi}\right|^2,
\end{equation}
where, in the last equality, we have used the expression $E^\text{scat}=E_z^1$ for the field scattered in the unbounded ($b_n^1=0$) medium~1 from Eq.~\eqref{eq:fields_expansion_E}, applied the far-field approximation $H_n^{(2)}(k_\ell\rho\rightarrow\infty)=\sqrt{2/(\pi k_\ell\rho)}e^{-\mathrm{i}(k_\ell\rho-n\pi/2-\pi/4)}$, and assumed that the incident electric field is a plane wave with unit magnitude, i.e, $E^\text{inc}=E^\text{ext}_{\text{i},z}=\text{e}^{-\text{j}k_0x}$.

The total scattering width, $\sigma$, which is the quantity to be minimized for cloaking, is then obtained upon integrating the echo width~\eqref{eq:SEW} over all the scattering angles as
\begin{equation}\label{eq:TSW}
\sigma=\frac{1}{2\pi}\int_0^{2\pi}\delta(\phi)d\phi=\frac{2}{\pi k_0}\int_0^{2\pi}\left|\sum_{n=-N}^{n=N}a_n^1\text{e}^{\text{j}n\phi}\right|^2d\phi.
\end{equation} 

For simplicity, we keep the innermost radius ($\rho_L=a$), the spacing between the metasurfaces ($d$), the number of metasurfaces ($L$), the wavenumber ($k_\ell$) and the wave impedance ($\eta_\ell$) fixed for all the $\ell$'s, and optimize only the bianisotropic susceptibilities tensors $\dya{\chi}^\ell$ ($\ell=1,2,\ldots L$). We perform this optimization iteratively, using an interior-point method, for the lowest normalized scattering width $\sigma_\text{norm}$, defined as the ratio of the total scattering width of the cloaked object~\eqref{eq:TSW} to that of the innermost (impenetrable) circular metasurface. We solve thus the optimization problem
\begin{equation}\label{eq:optimization}
\underset{\dya{\chi}^{\ell}}{\text{min}}~\sigma_\text{norm}\left(\dya{\chi}^{\ell},k_\ell,\eta_\ell,\rho_\ell,L\right),
\end{equation}
with a sufficiently large $L$ to obtain a sufficiently small minimum $\sigma_\text{norm}$ (e.g., $\sigma_\text{norm}^\text{min}=10^{-3}$), under the scattering constraint
{\begin{subequations}\label{eq:S_L_reciprocal}
\begin{equation}\label{eq:S11_mag}
\left|S_{11}^L\right|=1
\end{equation}
and
\begin{equation}
\left|S_{21}^L\right|=0,
\end{equation}
\end{subequations}
with $\angle S_{11}^L$ being a free design parameter. These constraints imply, according to~\eqref{eq:GSTC_susceptibility}, the following conditions on the fields at the two sides of the $L^\mathrm{th}$ metasurface:
\begin{subequations}\label{eq:int_ill_cond}
\begin{equation}
E_z^{+}=1+S^L_{11},\quad E_z^{-}=0
\end{equation}
and
\begin{equation}
H_\phi^{+}=\frac{1-S^L_{11}}{\eta_0},\quad H_\phi^{-}=0,
\end{equation}
\end{subequations}
with $S_{11}=\mathrm{e}^{j\angle S_{11}}$ according to~\eqref{eq:S11_mag}, where a specific value if found for $\angle S_{11}$ at the end of the optimization procedure.}

\subsection{\label{subsec:transmission}Transmission-Regime Beam Forming}

{The illumination assumption is now $\vec{E}^\text{ext}_\text{i}=0$ with $\vec{E}^\text{in}_\text{i}\neq 0$, and we impose the transmission constraints
\begin{subequations}\label{eq:S_L_nonreciprocal}
	\begin{equation}
	\left|S^L_{22}\right|=0
	\end{equation}
	and
	\begin{equation}
	\left|S_{12}^L\right|=1,
	\end{equation}
\end{subequations}
which are naturally nonreciprocal in conjunction with~\eqref{eq:S_L_reciprocal}, with  $\angle S_{12}^L$ being a free design parameter that we arbitrarily set to zero. These constraints imply, according to~\eqref{eq:GSTC_susceptibility}, the following conditions on the fields at the two sides of the $L^\mathrm{th}$ metasurface: 
\begin{subequations}\label{eq:ext_ill_cond}
	\begin{equation}
	E_z^{+}=1,~E_z^{-}=1,
	\end{equation}
	\begin{equation}
	H_\phi^{+}=-\frac{1}{\eta_0},~H_\phi^{-}=-\frac{1}{\eta_0}.
	\end{equation}
\end{subequations}}

{
\subsection{Susceptibility Parameters of the $L^\mathrm{th}$ Metasurface}
Separately inserting~\eqref{eq:int_ill_cond} (external illumination condition) and~\eqref{eq:ext_ill_cond} (internal illumination condition) into~\eqref{eq:GSTC_susceptibility}, and solving the resulting system of four equations for the susceptibility parameters yields
\begin{subequations}\label{eq:X_L_nonreciprocal}
	\begin{equation}
	\chi_{\textrm{ee}}^{zz,L}=\chi_{\textrm{em}}^{z\phi,L}=-\frac{\text{j}}{k_0}\left(1-S^L_{11}\right)
	\end{equation}
	and
	\begin{equation}
	\chi_{\textrm{me}}^{\phi z,L}=\chi_{\textrm{mm}}^{\phi\phi,L}=-\frac{\text{j}}{k_0}\left(1+S^L_{11}\right),
	\end{equation}
\end{subequations}
where we recall that $S_{11}=\mathrm{e}^{j\angle S_{11}}$ with $\angle S_{11}$ determined by the cloaking optimization (Sec.~\ref{subsec:cloaking}).}

{The relations~\eqref{eq:X_L_nonreciprocal} fully determine the $L^\mathrm{th}$ metasurface, not leaving out any \emph{metasurface} degrees of freedom, beyond the essential conditions~\eqref{eq:ext_ill_cond}, for transmission optimization. However, this is not excessively constraining because i)~the cloaking-optimized globally
transmissive nature of the metasurfaces $1$ to $L-1$ automatically provide an exit channel to the transmitted and ii)~the radiation characteristics of this wave may be independently
controlled using antenna design principle, as will be seen in Sec.~\ref{sec:results}.}


\section{\label{sec:results}Full-Wave Results}
%
We consider in this section a transmittable nonreciprocal cloak (Fig.~\ref{fig:design}) with a uniform metasurface spacing of $d=\lambda/4$ and a core radius of $\rho_L=\lambda$, where $\lambda$ is the wavelength of the waves to manipulate. Applying the design procedure outlined in Sec.~\ref{sec:design}, {with an interior-point optimization tool,} we found that $N=8$ metasurfaces are required to achieve $\sigma_\mathrm{norm}<10^{-3}$ under these conditions. The following presents the corresponding (full-wave) results, all of which are produced with the tools established in Sec.~\ref{sec:design}.

\begin{figure}[h!]
	\subfigure[]{
		\includegraphics[width=1\columnwidth]{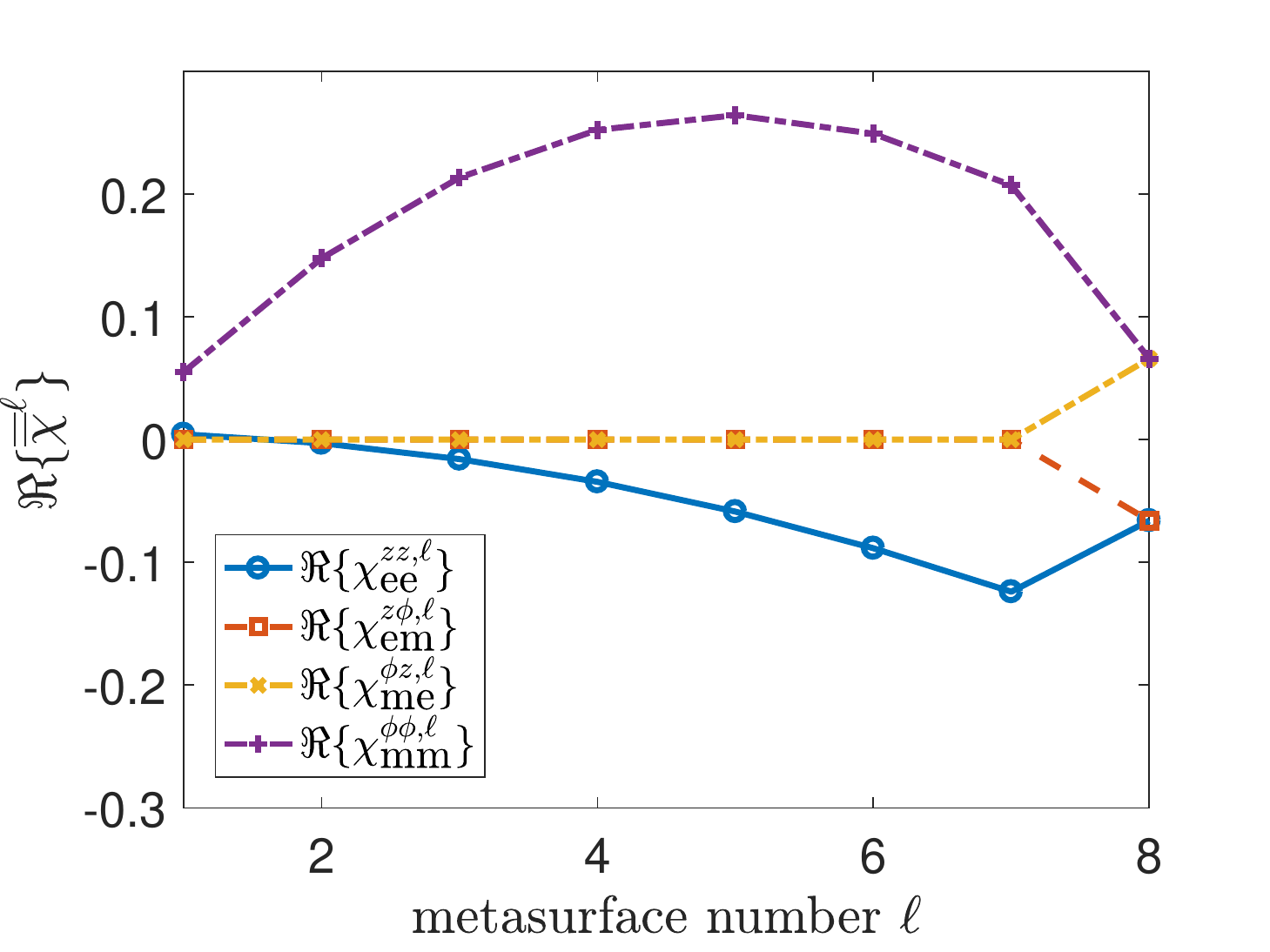}}
	\subfigure[]{
		\includegraphics[width=1\columnwidth]{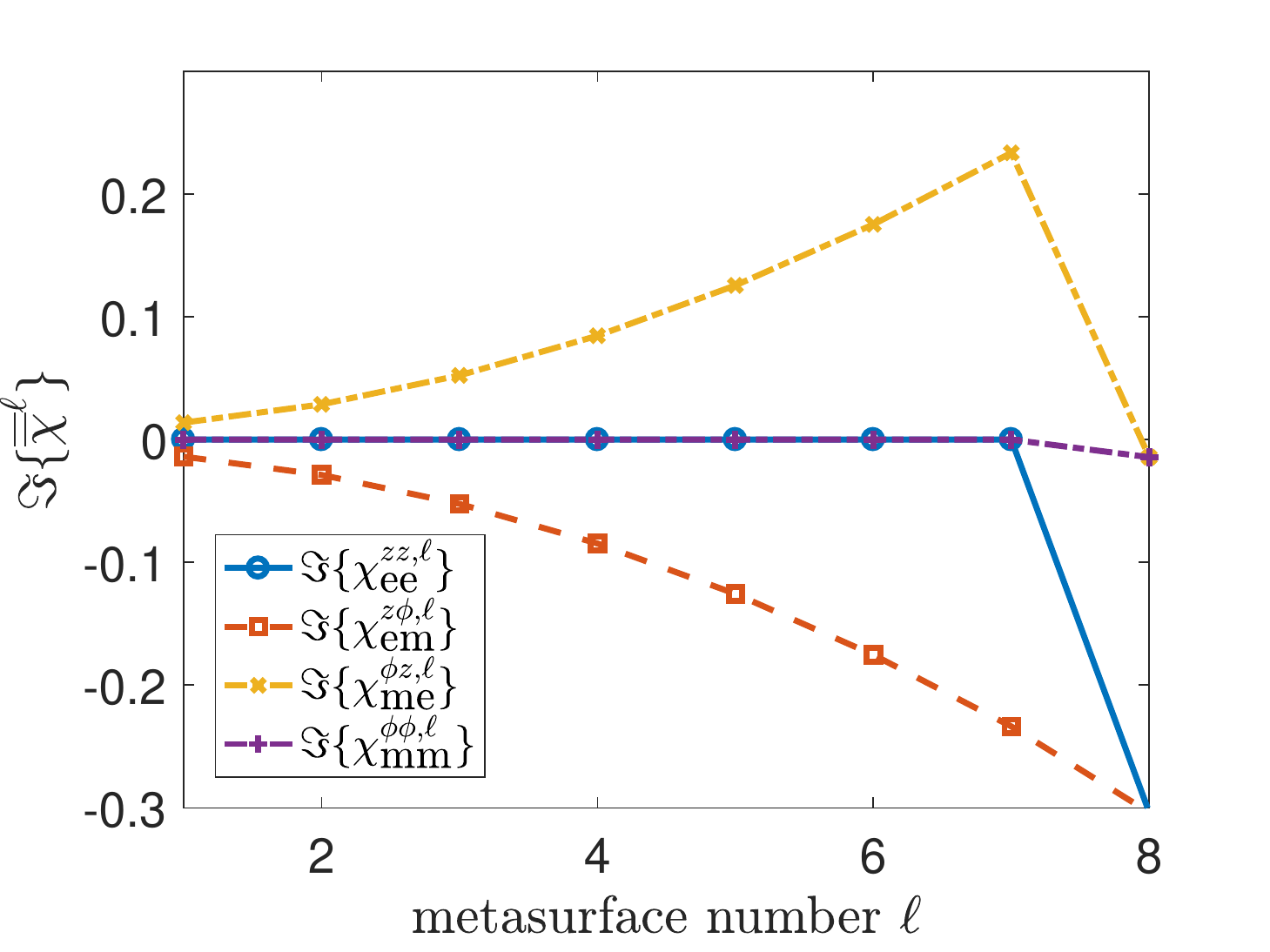}}
	\caption{\label{fig:results_susceptibilities}Susceptibilities of the metasurfaces for a transmittable nonreciprocal cloaking structure (Fig.~\ref{fig:design}) composed of $N=8$ metasurfaces with uniform spacing $d=\lambda/4$ and with core radius $\rho_L=\lambda$. (a)~Real parts. (b)~Imaginary parts.}
\end{figure}

Figure~\ref{fig:results_susceptibilities} shows the bianisotropic susceptibility parameters~\eqref{eq:susc_tens} {obtained by the cloaking optimization in Sec.~\ref{subsec:cloaking} for the metasurfaces $1$ to $L-1$ and by~\eqref{eq:X_L_nonreciprocal} for the metasurface $L$}, with the real and imaginary parts plotted in Figs.~\ref{fig:results_susceptibilities}(a) and~\ref{fig:results_susceptibilities}(b), respectively. Note that $\Re\{\chi^{z\phi,\ell}_\text{em}\}=\Re\{\chi^{\phi z,\ell}_\text{me}\}=\Im\{\chi^{zz,\ell}_\text{ee}\}=\Im\{\chi^{\phi \phi,\ell}_\text{mm}\}=0$ and $\chi^{z\phi,\ell}_\text{em}=\chi^{\phi z,\ell}_\text{me}$ for $\ell=1,2,\ldots,7$, according to the lossless-gainless and reciprocity specifications, respectively, whereas the these conditions are broken in the $\ell=8^\mathrm{th}$ metasurface, according to the nonreciprocity specification and related lossy condition~\cite{Caloz_PRAp_10_2018}. The curves in Fig.~\ref{fig:results_susceptibilities} exhibit an overall trend of parameter increasing in magnitude from the outer to the inner layers of the system, as intuitively expected from the fact that deeper layers require stronger wave deviation for cloaking.

\begin{figure}[h!]
	\subfigure[]{
		\includegraphics[width=1\columnwidth]{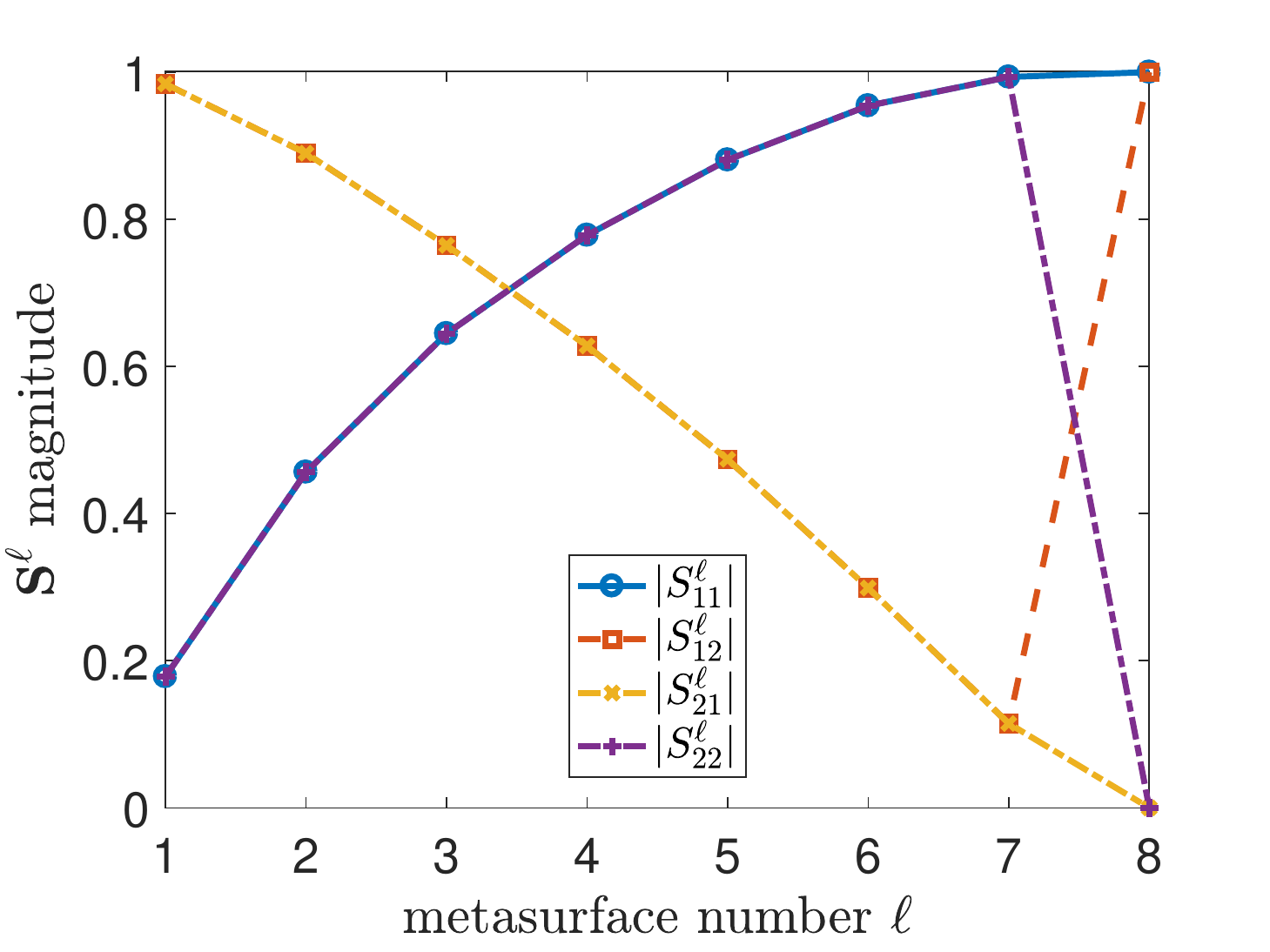}}
	\subfigure[]{
		\includegraphics[width=1\columnwidth]{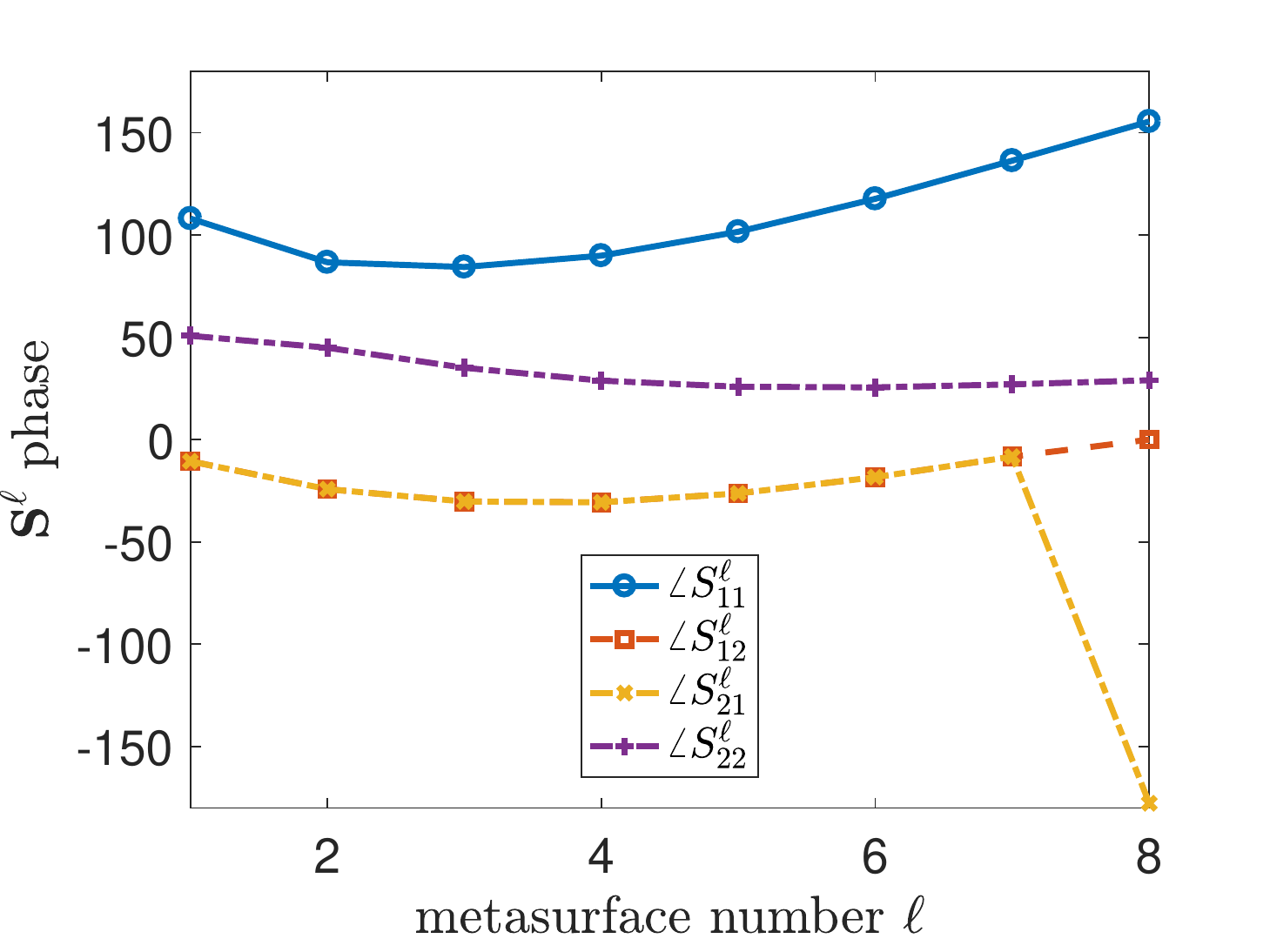}}
	\caption{\label{fig:results_Sparameters}Scattering parameters corresponding to the susceptibilities in Fig.~\ref{fig:results_susceptibilities} under normal incidence. (a)~Magnitude. (b)~Phase.}
\end{figure}

Figure~\ref{fig:results_Sparameters} shows the scattering parameters corresponding to the susceptibilities in Fig.~\ref{fig:results_susceptibilities} under normal incidence, {computed via conversion formulas provided in~\cite{achouri2021electromagnetic},}, with the magnitude and phase parts plotted in Figs.~\ref{fig:results_Sparameters}(a) and~\ref{fig:results_Sparameters}(b), respectively.

Consistently with Fig.~\ref{fig:results_Sparameters}(a), the metasurfaces become progressively more reflective in the outer to inner direction of the structure with the last metasurface being totally reflective and opaque from the exterior (i.e., $\left|S_{11}^L\right|=1$ and $\left|S_{21}^L\right|=0$) and perfectly matched and transmissive from the interior (i.e., $\left|S_{22}^L\right|=0$ and $\left|S_{12}^L\right|=1$). The result $\angle S^L_{11}\approx 155^\circ$ indicates that the innermost metasurface exhibits an external response that is fairly close but not exactly equal to that of a perfect electric conductor ($\angle S_{11}=180^\circ$).

{All the forthcoming results, until the end of the paper, have been obtained using the electromagnetic analysis presented in Sec.~\ref{subsec:analysis}, based on optimized susceptibility results of the type in Fig.~\ref{fig:results_susceptibilities}.}

\subsection{\label{subsec:external_illumination}External Illumination}
\begin{figure*}
    \subfigure[]{
    	\includegraphics[width=1\columnwidth]{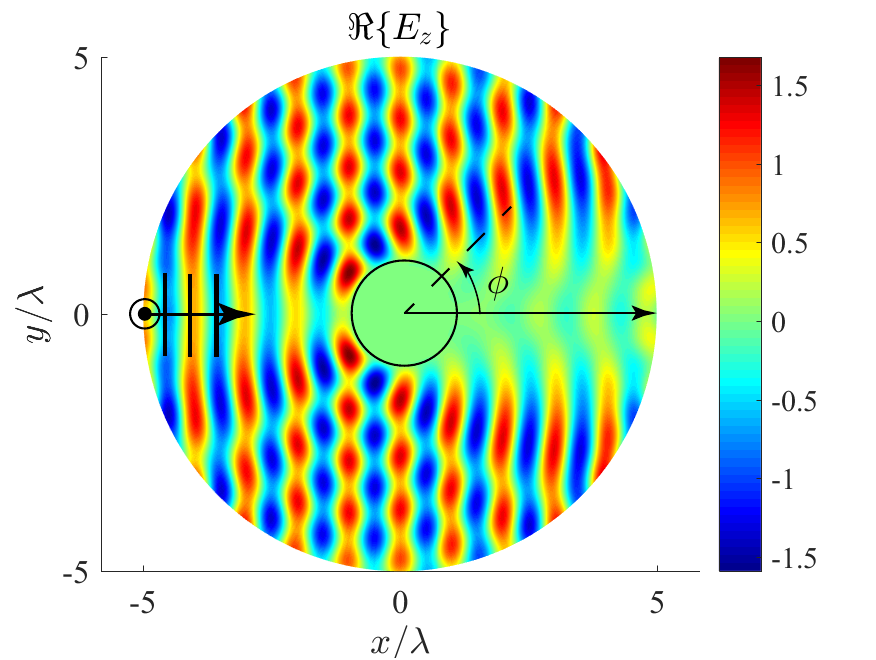}}
	\subfigure[]{
    	\includegraphics[width=1\columnwidth]{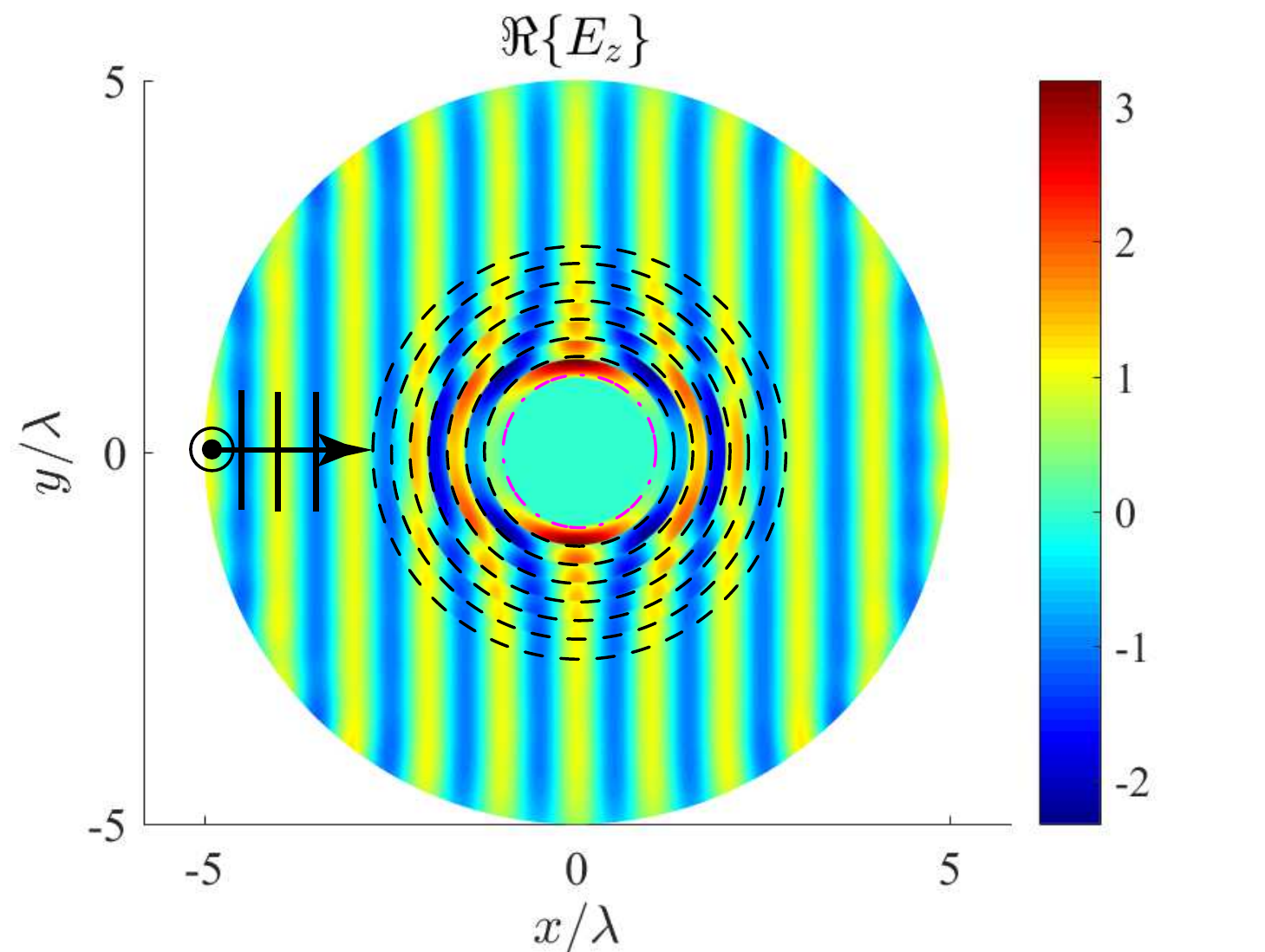}}
    \subfigure[]{
    	\includegraphics[width=1\columnwidth]{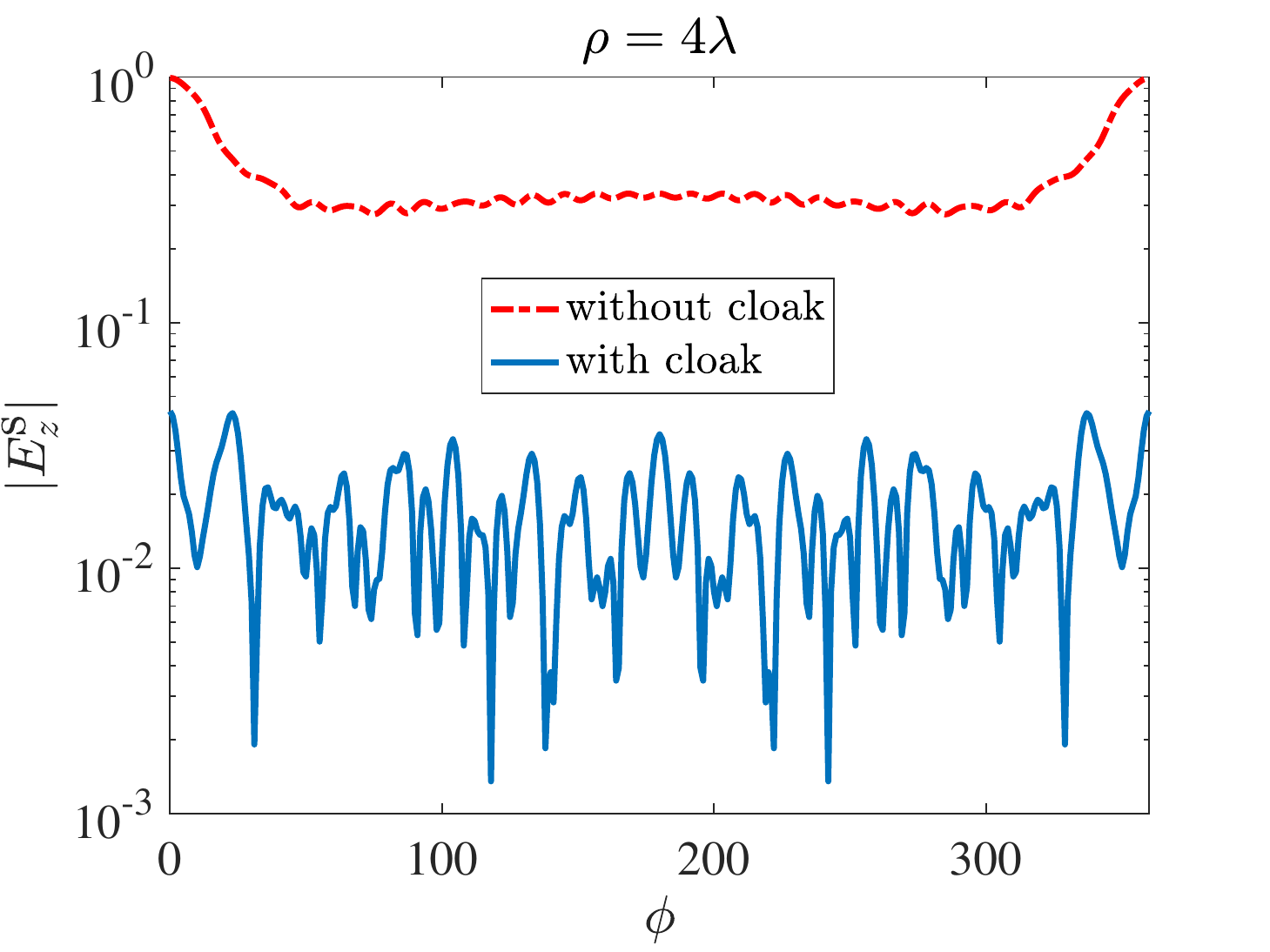}}
    \subfigure[]{
        \includegraphics[width=1\columnwidth]{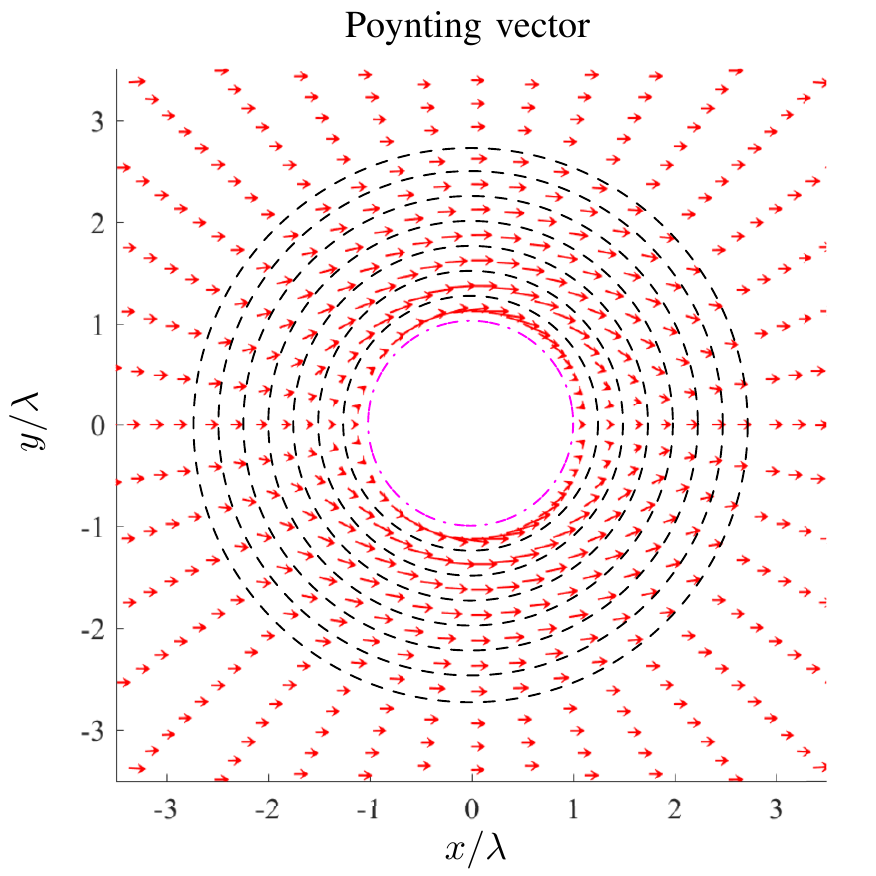}}
	\caption{\label{fig:cloak}Cloaking under (external) plane-wave illumination for the design in Figs.~\ref{fig:results_susceptibilities} and~\ref{fig:results_Sparameters}. {Real part of the total electric field for (a)~an impenetrable object without cloaking and (b)~the same object surrounded by the proposed cloak. (c)~Magnitude of the scattered field in (a) and (b) in the circular section $\rho=4\lambda$. (d)~Poynting vector corresponding to (b).}}
\end{figure*}

%
Figure~\ref{fig:cloak} presents the cloaking result under (external) plane-wave illumination, with {Figs.~\ref{fig:cloak}(a) to~(d) plotting the response of an impenetrable object without cloak, for comparison, the response of the same object surrounded by the proposed cloak, the comparative responses of the two previous structures in a circular section of space, and the Poynting vector field corresponding to Fig.~\ref{fig:cloak}(b), respectively.} Quasi-perfect cloaking is observed. Note that the Poynting vector provides an insightful perspective of the multiple-scattering deviation mechanism in the concentric metasurface cloak structure, which, as may have been intuitively expected, is not so different from that of the coordinate-transformation deviation. Although the cloaking result is shown here for one angle, the device is circularly symmetric structure, as previously mentioned, and exhibits hence exactly the same cloaking performance for any incidence angle (omnidirectional cloaking).

\begin{figure}[h!]
	\includegraphics[width=1\columnwidth]{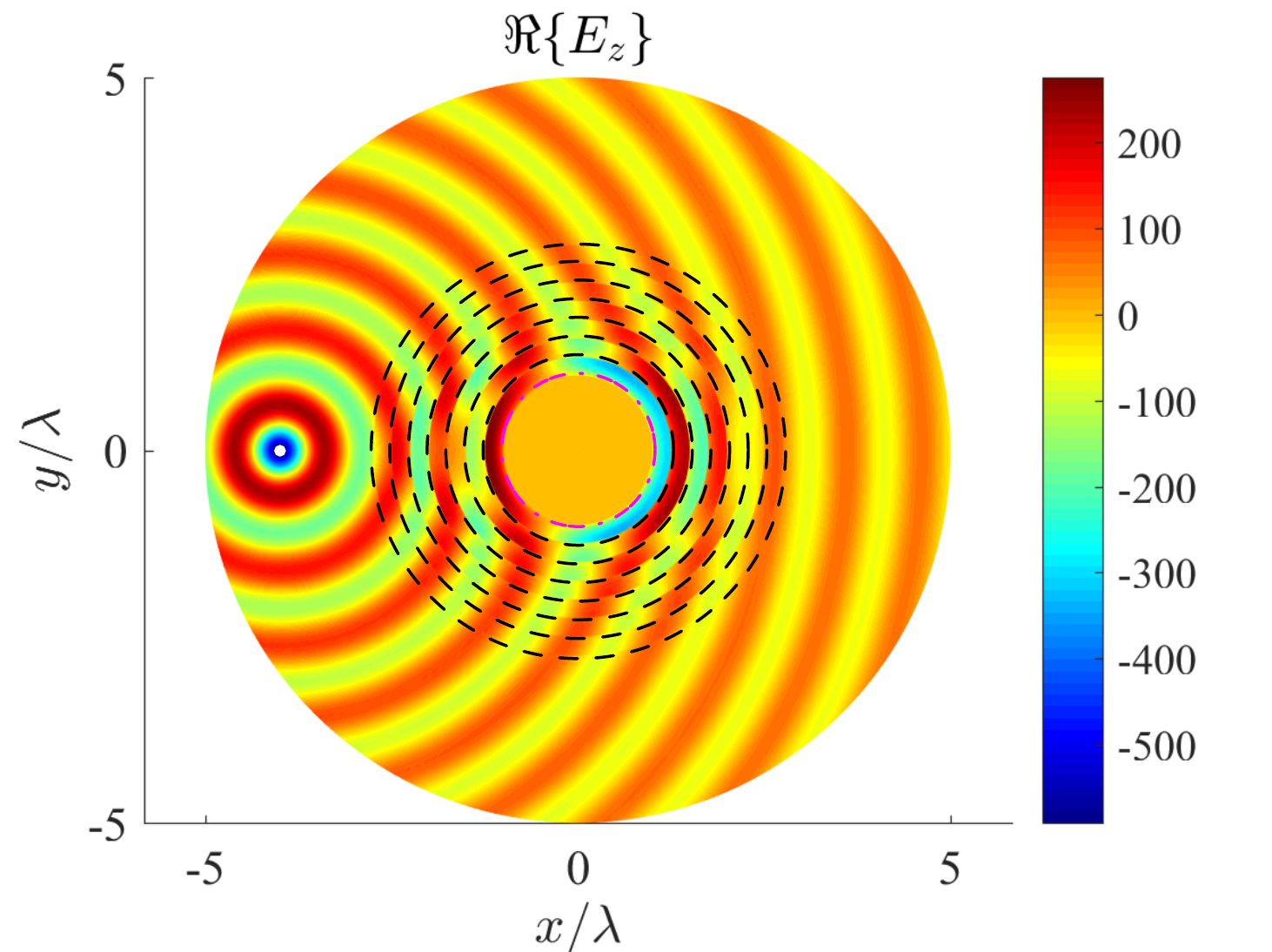}
	\caption{\label{fig:ext_point_source}Cloaking under (external) point/line-source illumination for the design in Figs.~\ref{fig:results_susceptibilities} and~\ref{fig:results_Sparameters} (real part of the total electric field), source placed at the point $\rho=4\lambda$ and $\phi=\pi$ (see Fig.~\ref{fig:design}).}
\end{figure}

%
Figure~\ref{fig:ext_point_source} presents the cloaking result under (external) point/line-source illumination. Here, again, a quasi-perfect cloaking result is observed. This  insensitivity of the structure to the nature of the source in cloaking may a priori seem surprising given that the cloak design is based on optimization under plane-wave incidence and not on a fundamental, angle-independent scheme such as the coordinate-transformation one. The reason is that, although the plane wave impinges normally ($\phi=0^\circ$) onto the equator of the cloak, it impinges on the latitudes from the equator to the poles with a continuum of all possible incidence angles ($\phi=0^\circ\rightarrow 90^\circ$). Therefore, the optimization process automatically accounts for all the directions included in the angular spectrum of the point source (or any other source), and the cloak is hence working for any source topology.

\subsection{\label{subsec:internal_illumination}Internal Illumination}
\begin{figure}[h!]
    \includegraphics[width=1\columnwidth]{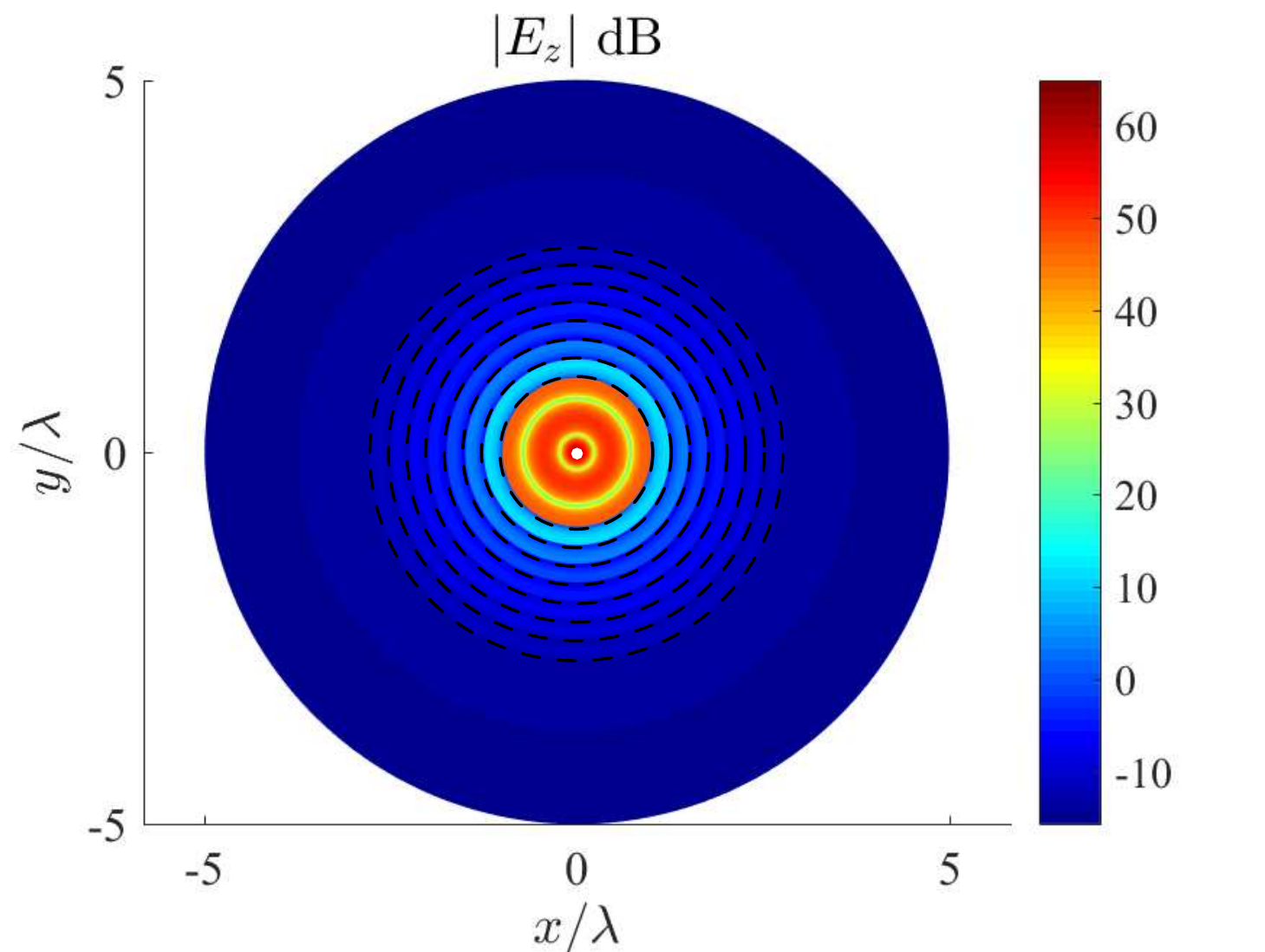}
	\caption{\label{fig:internal}Response of the structure with the design in Figs.~\ref{fig:results_susceptibilities} and~\ref{fig:results_Sparameters} (real part of the total electric field) under internal illumination (point source at the center) with nonreciprocity turned off.}
\end{figure}

%
As pointed out in Sec.~\ref{sec:principle} and illustrated in Fig.~\ref{fig:concept}(a), in a properly designed reciprocal cloak, light launched from the core of the device should be essentially reflected back by the cloak shell. Figure~\ref{fig:internal}, displaying the response of the system under internal illumination with nonreciprocity turned off, shows that the selected concentric metasurface cloaking technique exhibits indeed this characteristic in the absence nonreciprocity. The light confinement in the core is not perfectly clear, with the negligible leakage beyond the cloak shell ($\sim50$~dB below the average core field) due to the imperfectness of the design ($\sigma_\mathrm{norm}\approx 10^{-3}\neq 0$).

\begin{figure}[h!]
	\subfigure[]{
		\includegraphics[width=1\columnwidth]{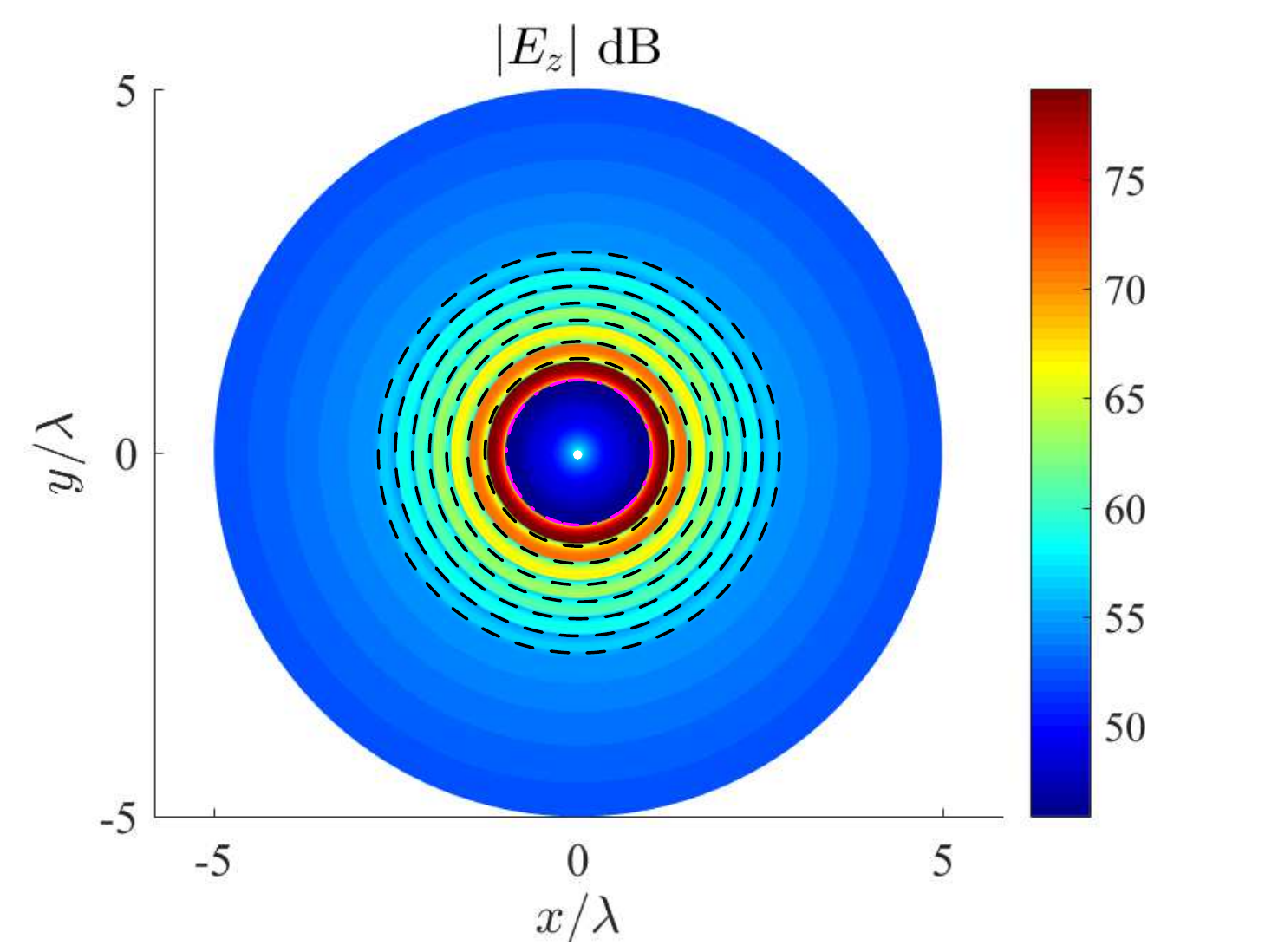}} \\
	\subfigure[]{
		\includegraphics[width=1\columnwidth]{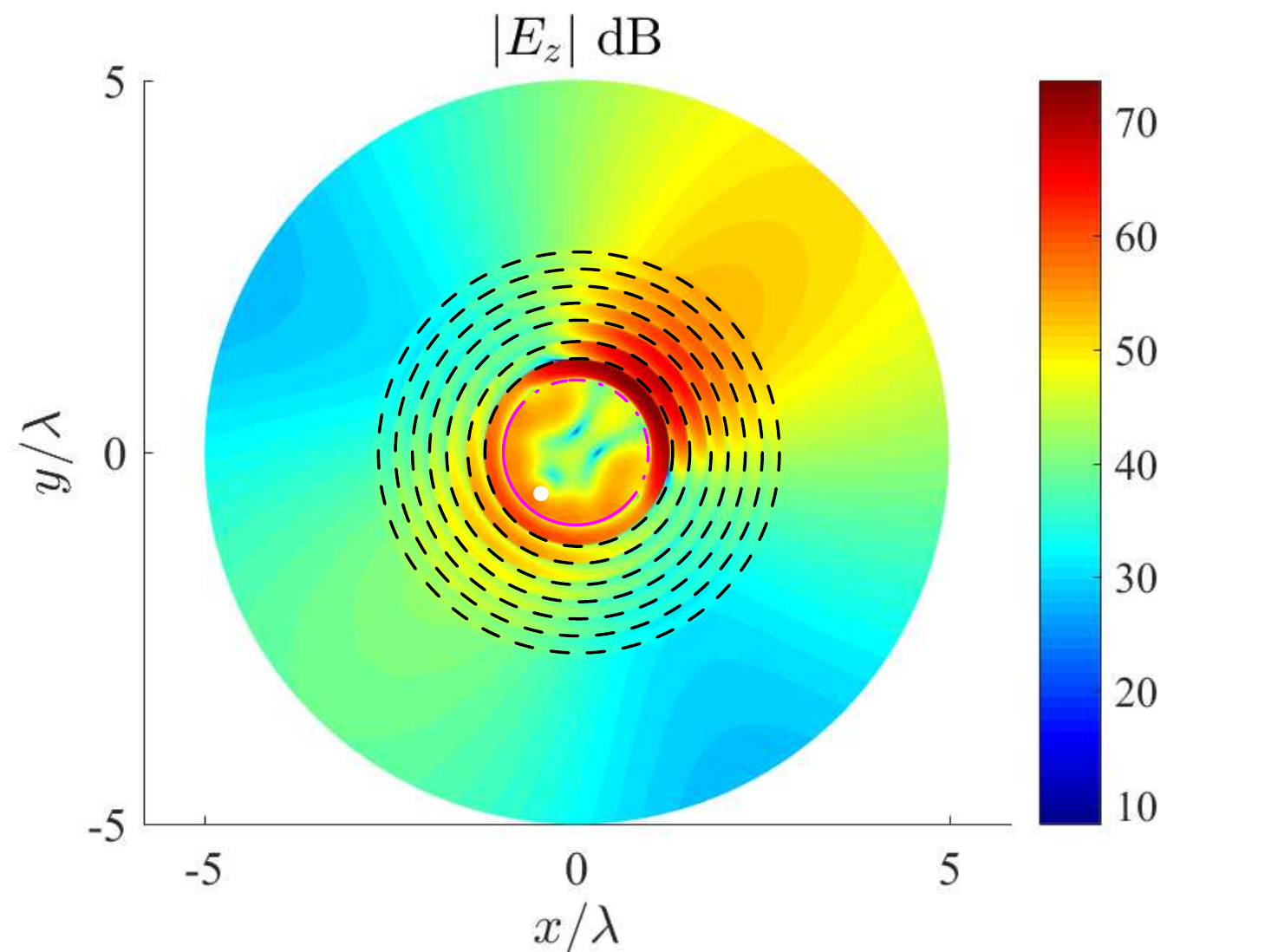}}
	\caption{\label{fig:transmission}Transmission under internal illumination for the design in Figs.~\ref{fig:results_susceptibilities} and~\ref{fig:results_Sparameters} (magnitude of the total electric field) with nonreciprocity turned on. (a)~Omnidirectional (point) source placed at the center of the structure. (b)~Directive radiation towards $\alpha=45^\textrm{o}$, with the point source placed at $(\rho^\prime,\phi^\prime)=(3\lambda/4,5\pi/4)$ backed by a half-circular reflector.}
\end{figure}

Finally, Fig.~\ref{fig:transmission} demonstrates the unique outward transmission capability of the proposed nonreciprocal cloak, with Figs.~\ref{fig:transmission}(a) and~\ref{fig:transmission}(b) showing omnidirectional transmission {from a centered isolated source} and directional transmission (with directivity of $7.67$~dB) {from an off-set mirror-backed source}, respectively. {As anticipated in Sec.~\ref{sec:principle}, the globally transmissive cloaking-optimized structure beyond the innermost (nonreciprocal) metasurface provides a proper exit channel to the wave originating from the core of the cloak. In fact, more sophisticated designs, still independent from the cloaking design or co-designed with it, could be achieved, such as higher-directivity radiation and advanced beam forming, using an array of a few antenna elements following standard antenna design techniques~\cite{balanis2015antenna}.}

{\subsection{\label{subsec:external_internal_illuminations}Simultaneous External and Internal Illuminations}}

\begin{figure}[h!]
	\subfigure[]{
		\includegraphics[width=1\columnwidth]{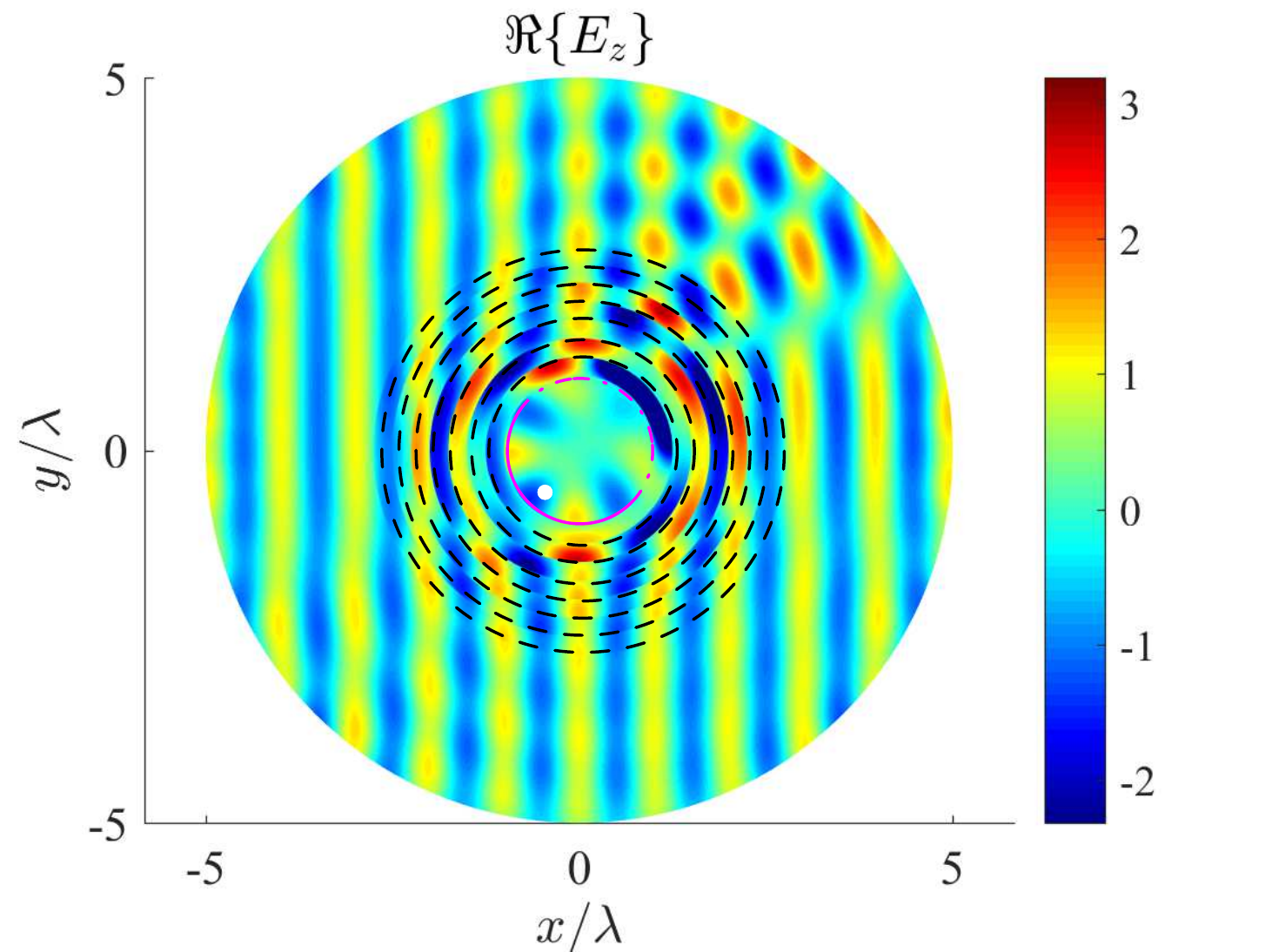}}
	\subfigure[]{
		\includegraphics[width=1\columnwidth]{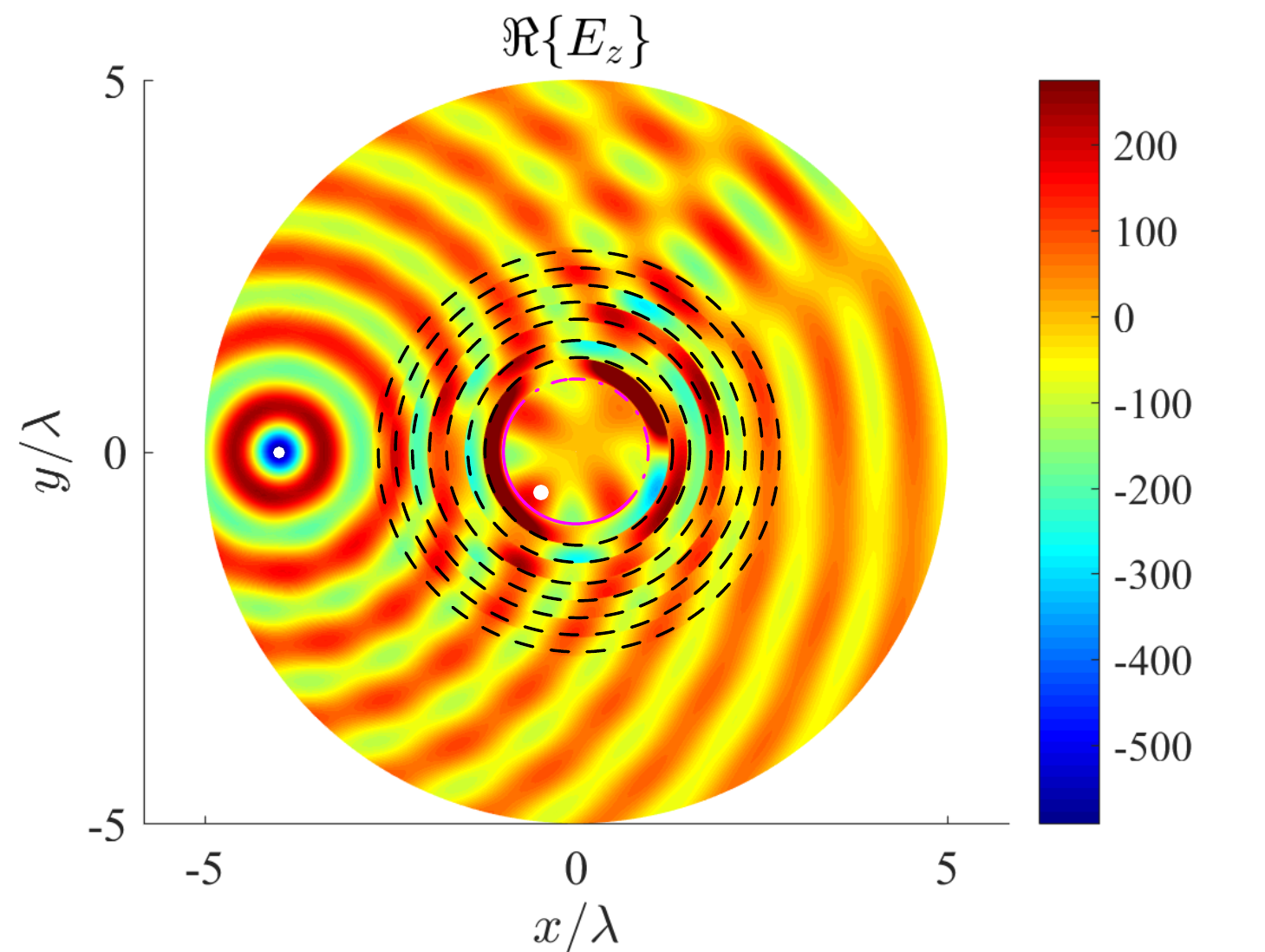}}
	\caption{\label{fig:simultaneous_excitations}{Response of the transmittable nonreciprocal cloak under simultaneous external and internal illuminations (real part of the total electric field) for internal point/line source located at $(3\lambda/4,5\pi/4)$ and (a)~external plane-wave source with wave amplitude unity along with internal source of $0.002$~A, and (b)~external point/line source strength of 1~A at $(4\lambda,\pi)$ along with internal point/line source strength of $-0.25$~A.}}
\end{figure}

{The proposed transmittable nonreciprocal cloak may operate either in ``simplex mode'', whereby either only the cloaking operation [$\vec{E}^\text{ext}_\text{i}\neq 0$ but $\vec{E}^\text{int}_\text{i}=0$ (silent transmitter)] or the transmitting function [$\vec{E}^\text{int}_\text{i}\neq 0$ but $\vec{E}^\text{ext}_\text{i}=0$ (non-illuminated cloak)] is active at a given time, as illustrated in Figs.~\ref{fig:cloak}(b) and~\ref{fig:ext_point_source} for the former case, and Fig.~\ref{fig:transmission} for the latter case. However, it is especially conceived to operate in ``full-duplex mode'' [$\vec{E}^\text{ext}_\text{i}\neq 0$ and $\vec{E}^\text{int}_\text{i}\neq 0$], where the two operations are simultaneously performed. This scenario is illustrated in Fig.~\ref{fig:simultaneous_excitations}. This figure, which naturally corresponds to a superposition of the separate illumination results given the linearity of the overall system, provides a visual sense of the duplex operation of the system, whereby transmission is effectively produced in the intended direction while cloaking is realized everywhere else. Note that information carried by the transmitted wave could be easily received, despite the presence of the external source, using proper communication modulation techniques~\cite{lathi1995modern}.}

\begin{figure}[h!]
	\includegraphics[width=1\columnwidth]{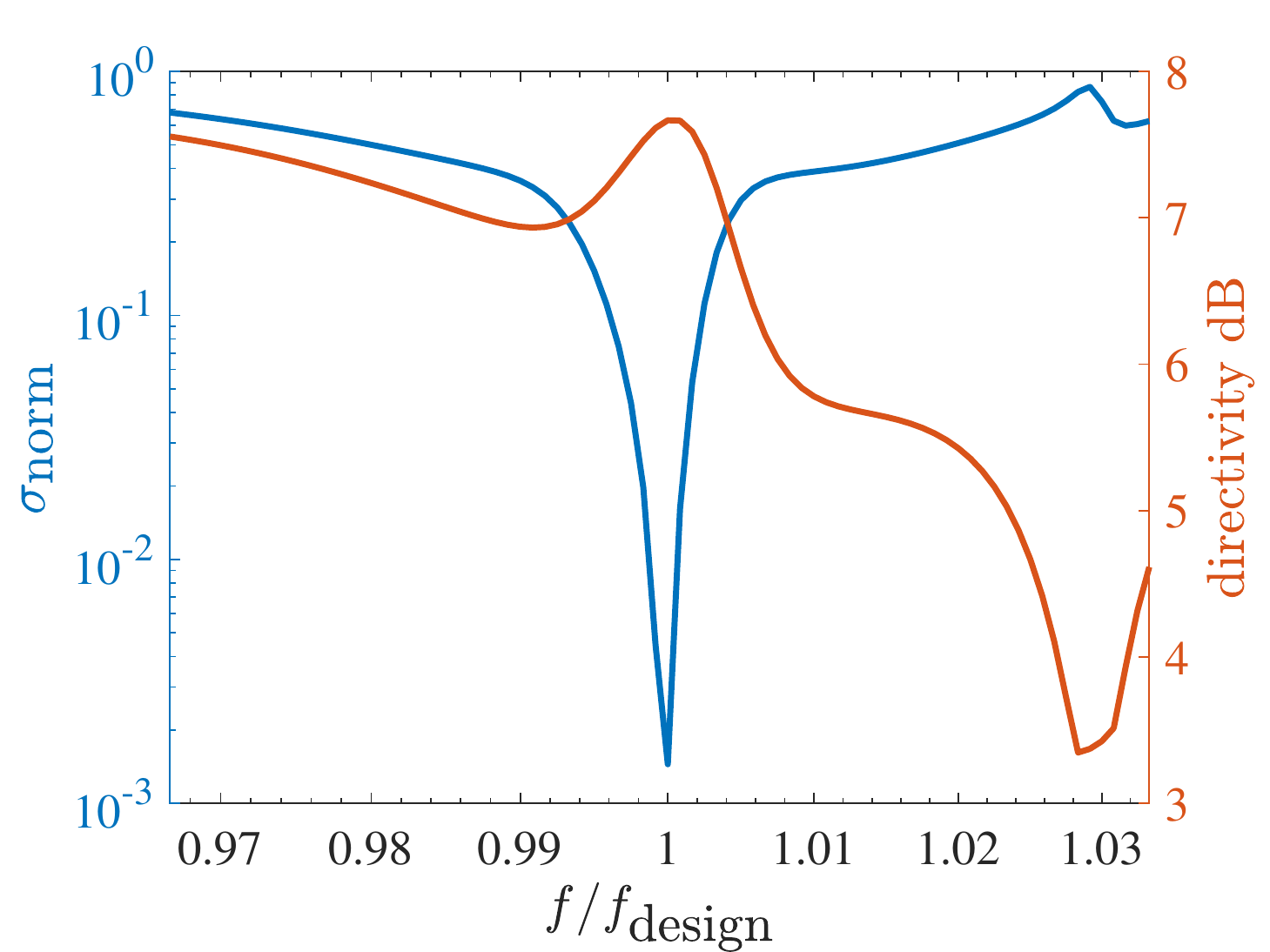}
	\caption{\label{fig:frequency_response}{Frequency response of the nonreciprocal transmittable cloak in terms of (normalized) scattering width [Eq.~\eqref{eq:TSW}] for external illumination (cloaking), with the parameters in Fig.~\ref{fig:cloak}(b), and directivity for internal illumination (transmission), with the parameters in Fig.~\ref{fig:transmission}(b).}}
\end{figure}

{\subsection{\label{subsec:frequency_response}Bandwidth Considerations}}
{Figure~\ref{fig:frequency_response} plots the frequency response of the nonreciprocal transmittable cloak in both the cloaking and transmission regimes. The bandwidth of the system involves two aspects: i)~the bandwidth of the (Lorentz-type) resonant particles forming the metasurfaces and ii)~the bandwidth of the overall circular Fabry-Perot resonator structure assuming unlimited-bandwidth metasurface particles. The bandwidth of the latter is bounded by the Fabry-Perot etalon layer having the most reflective interfaces, since the bandwidth of a Fabry-Perot etalon is inversely proportional to the product of its interface reflectances [$\text{BW}=2/\mathcal{F}$, with finesse $\mathcal{F}=\pi\sqrt{|r_1 r_2|}/(1-|r_1r_2|)$]~{\cite{saleh2019fundamentals}}. This typically corresponds to the innermost layer of the cloaking structure, as illustrated in {Fig.~\ref{fig:results_Sparameters}}(a); in this design, the reflectance product is of $0.99$, which yields a bandwidth of about $0.6\%$, consistently with the bandwidth of both the cloaking and transmission curves in {Fig.~\ref{fig:frequency_response}}. On the other hand, the bandwidth of resonant particles, which essentially depends on their specific geometries, is typically in the order of $5\%$~{\cite{achouri2021electromagnetic}}; this is one order of magnitude larger than the aforementioned Fabry-Perot resonance, and the metaparticle bandwidth limitation does therefore not impact the bandwidth of the overall system.
%
}     


\section{\label{sec:potential}Potential Applications}

This section describes some of the potential applications of the proposed transmittable nonreciprocal cloaking (Fig.~\ref{fig:structure}).

\subsection{Selective Cloaking}\label{subsec:stealth}
As any cloaking system, the proposed device may be used for camouflaging from radar, since the interrogating wave is deviated by the cloak around the object placed in its core without any reflection, as well as without scattering that could be detected by another foe  placed somewhere else. But this device offers the extra functionality of \emph{camouflaging selectivity} whereby the host of the cloak can communicate with friends while being undetectable by foes. This may be accomplished in two fashions. If the position of the friend is known, one may use the directional option in Fig.~\ref{fig:transmission}(b), and possibly even rotate (mechanically or electronically) the antenna system to follow this friend or reach other friends in different directions. If the position of the friend(s) is unknown, one may use the omnidirectional mode in Fig.~\ref{fig:transmission}(a) along with a spread-spectrum encryption key, so that the transmitted signal is spread out to a level below the noise floor of the foe, but can be restored by the friend(s) upon multiplication with the encryption key in their possession~\cite{wang2021Camouflaging}.

\subsection{Blockage Avoidance}\label{subsec:blockage} 
The feed of a parabolic antenna should ideally be dimensionless to avoid perturbing, by blockage and diffracting, the waves reflected by the parabolic dish. Unfortunately, the feed must have a size that is comparable to the wavelength for efficient radiation and even substantially larger than it when over-spilling constraints require high feed directivity. The proposed device can resolve this issue in the antenna transmission mode. Indeed, placing the feed in the core of the nonreciprocal (omnidirectional) cloak allows the signal to radiate across the cloak so as illuminate the dish while the wave subsequently reflected by the dish are deviated around the feed via cloaking. This scheme would not readily work in the antenna receiving mode, where the received signal would be appropriately deviated by cloaking around the feed, but could not then penetrate inside the cloak to reach the feed (2 equivalent external sources); in that case, a simple solution would be to use a half (reciprocal) cloak with the cloak side of course oriented toward the incidence side and the dish side being left empty.

\subsection{Electromagnetic Illusion}\label{subsec:illusion}
Electromagnetic (or optical) illusion has been mostly realized by the transformation-coordinate technique so far~\cite{lebbe2022susceptibility}. The proposed metasurface-based cloaking approach, given its greater fabrication simplicity and bianisotropic (36 accessible parameters) flexibility~\cite{achouri2021electromagnetic}, has a potential for more diverse illusion operations. Moreover, the proposed nonreciprocity functionality could further enrich the illusion efficacy by having the entity in the core of the cloak launching strong deceptive signals, possibly with elaborate space-time spectral transformations~\cite{Chamanara_TAP_04_2019}.

\subsection{Cooling Window}\label{subsec:cooling}
Significant efforts have been realized in recent years to realize smart windows that optimize thermal radiation in order to save energy~\cite{Zhou2021cooling}. An ideal window of that type would, for instance in the summer for saving cooling energy, transmit indoors heat outwards while reflecting outdoors (solar and environmental) heat, a clearly nonreciprocal operation, that would benefit from nonreciprocal metasurfaces~\cite{Lavigne_NJP_07_2021,lavigne2022metasurface} operating at the appropriate infrared and far-infrared wavelengths. In this area, the additional cloaking feature of the proposed device, which might be implemented in windows of various (curved) shapes may offer further thermal control flexibility in near future.


\section{\label{sec:conclusion}Conclusion}

We have introduced the concept of transmittable nonreciprocal cloaking and demonstrated it a concentric metasurface structure. This metasurface represents a fundamental diversification of the already powerful concept of cloaking and has potential application, some of which have been described in the last part section of the paper.

\section*{Supplementary Material}

\subsection{\label{appendix:model}Metasurface Modeling}

Assuming, for simplicity, a metasurface involving only \emph{tangential} electric and magnetic surface polarization densities, the GSTCs read to~\cite{dehmollaian2019comparison} [Fig.~\ref{fig:structure}(b)]
\begin{subequations}\label{eq:GSTC}
	\begin{equation}
	\hat \rho \times \Delta \vec E = -\text{j}{k_0}{\eta_0}{{\vec M}_{\text{s},\|}},
	\end{equation}
	\begin{equation}
	\hat \rho \times \Delta \vec H =  \text{j}\omega {{\vec P}_{\text{s},\|}},
	\end{equation}
\end{subequations}
where $\hat \rho$ is the unit vector normal to the surface of the metasurface, $\Delta$ refers to the difference of the fields (electric $\vec E$ and magnetic $\vec H$) at both sides of the metasurface at $\rho=\rho_\ell^-$ and $\rho=\rho_\ell^+$ (e.g., $\Delta \vec E=\vec E^{+} - \vec E^{-}$), $k_0$ and $\eta_0$ are the free-space wavenumber and wave impedance, respectively, $\omega$ is the angular frequency, and $\vec{M}_{\text{s},\|}$ (A) and $\vec{P}_{\text{s},\|}$(C/m) are the tangential surface magnetic and electric polarization densities, respectively, with the symbol $\|$ denoting vector components tangential to the metasurface. 

In this model, $\vec{M}_{\text{s},\|}$ and $\vec{P}_{\text{s},\|}$ are expressed in terms of the susceptibility tensors $\dya{\chi}_{\text{ee}\|}$, $\dya{\chi}_{\text{em}\|}$, $\dya{\chi}_{\text{me}\|}$ and $\dya{\chi}_{\text{mm}\|}$, 
and of the average electric and magnetic fields at the metasurface sheet [i.e., $\vec E_{\text{av}}=(\vec E^{+} + \vec E^{-})/2$ and $\vec H_{\text{av}}=(\vec H^{+} + \vec H^{-})/2$), i.e.,
\begin{subequations}\label{eq:MP}
	\begin{equation}
	{\vec M}_{\text{s},\|}
	={\frac{1}{\eta_0}} \overline {\overline{\chi}} _{\text{me}\|} \cdot \vec E_{\|,\,\text{av}} + \overline{\overline{\chi}} _{\text{mm}\|} \cdot \vec H_{\|,\,\text{av}},
	\end{equation}
	\begin{equation}
	{\vec P}_{\text{s},\|} 
	={\epsilon _0} \overline {\overline{\chi}} _{\text{ee}\|} \cdot \vec E_{\|,\,\text{av}} + \frac{1}{\text{c}} \overline{\overline{\chi}} _{\text{em}\|} \cdot \vec H_{\|,\,\text{av}},
	\end{equation}
\end{subequations}
where ${\epsilon _0}$ and $\text{c}$ are the free-space permittivity and speed of light, respectively. In the problem at hand (see Fig.~\ref{fig:structure}), where the fields are s-polarized, the only contributing susceptibilities in~\eqref{eq:MP} are  
\begin{subequations}\label{eq:susc}
	\begin{equation}
	\dya{\chi}_{\text{ee}\|}=\chi_\text{ee}^{zz}~\hat{z}\hat{z},
	\end{equation}
	\begin{equation}
	\dya{\chi}_{\text{em}\|}=\chi_\text{em}^{z\phi}~\hat{z}\hat{\phi},
	\end{equation}
	\begin{equation}
	\dya{\chi}_{\text{me}\|}=\chi_\text{me}^{\phi z}~\hat{\phi}\hat{z},
	\end{equation}
	\begin{equation}
	\dya{\chi}_{\text{mm}\|}=\chi_\text{mm}^{\phi \phi}~\hat{\phi}\hat{\phi},
	\end{equation}
\end{subequations}

The GSTCs~\eqref{eq:GSTC_susceptibility} are then obtained by substituting~\eqref{eq:susc} into \eqref{eq:MP}, and then inserting the resulting equations into~\eqref{eq:GSTC}.

\subsection{\label{appendix:analysis}Scattering Analysis}
%
First, we expand the tangential fields in each layer $\ell$, $E_z^\ell$ and $H_\phi^\ell$, over the natural modes of the system, which are Bessel functions in the radial direction, $\rho$, multiplied by the complex exponential function in the azimuthal direction, $\phi$, namely
\begin{subequations}\label{eq:fields_expansion_appendix}
	\begin{align}
	E_z^\ell &= \sum_{n=-N}^{n=N}\text{j}^{-n}\left[b_n^\ell J_n(k_\ell \rho)+a_n^\ell H_n^{(2)}(k_\ell \rho)\right]\text{e}^{\text{j}n\phi}\nonumber\\
	&=\sum_{n=-N}^{n=N}\tilde{E}_n^\ell(\rho)~\text{e}^{\text{j}n\phi}
	\end{align}
	and
	\begin{equation}
	H_\phi^\ell=\frac{1}{\text{j}k_\ell\eta_\ell}\frac{\partial E_z^\ell}{\partial \rho}=\sum_{n=-N}^{n=N}\tilde{H}_n^{\ell}(\rho)~\text{e}^{\text{j}n\phi},
	\end{equation}
\end{subequations}
where we have written the expansions in the convenient form of Fourier series, whose spectral coefficients, $\tilde{E}_n^\ell$ and $\tilde{H}_n^\ell$, depend on $\rho$, and explicitly read
\begin{subequations}\label{eq:expansion_coefficients}
	\begin{equation}
	\tilde{E}_n^\ell(\rho) = \text{j}^{-n}\left[b_n^\ell J_n(k_\ell \rho)+a_n^\ell H_n^{(2)}(k_\ell \rho)\right]
	\end{equation}
	and
	\begin{equation}
	\tilde{H}_n^{\ell}(\rho) = \frac{\text{j}^{-(n+1)}}{\eta_\ell}\left[b_n^\ell J_n^\prime(k_\ell \rho)+a_n^\ell H_n^{(2)\prime}(k_\ell \rho)\right],
	\end{equation}
\end{subequations}

Then, we express the susceptibility parameters of each metasurface in terms of Fourier series, for later matching with the fields, i.e., 
\begin{subequations}\label{eq:susceptibility_FS}
	\begin{equation}
	\chi_{pq}^\ell(\phi) = \sum_{n=-2N}^{n=2N}\tilde{\chi}^\ell_{ab,n}~\text{e}^{\text{j}n\phi},
	\end{equation}
    with the spectral coefficients
	\begin{equation}
	\tilde{\chi}_{pq,n}^\ell = \frac{1}{2\pi}\int_0^{2\pi}{\chi}_{ab}^\ell(\phi)~\text{e}^{-\text{j}n\phi}d\phi,
	\end{equation}
\end{subequations}
where $p,q=\text{e,m}$, corresponding to the four nonzero susceptibilities ${{\chi}}_{\text{ee}}^{zz,\ell}$, ${{\chi}}_{\text{em}}^{z\phi,\ell}$, ${{\chi}}_{\text{me}}^{\phi z,\ell}$ and ${{\chi}}_{\text{mm}}^{\phi\phi,\ell}$, and where we dropped the superscripts $zz$, $z\phi$, $\phi z$ and $\phi \phi$ for conciseness.

Finally, we apply the mode matching technique at each metasurface boundary, $\rho = \rho_\ell$, in Fig.~\ref{fig:design}, by inserting~\eqref{eq:fields_expansion_appendix} and~\eqref{eq:susceptibility_FS} into~\eqref{eq:GSTC_susceptibility}. This yields 
\begin{subequations}\label{eq:GSTC_spectral}
	\begin{equation}
	\Delta \tilde{E}_n^\ell = \text{j}k_0\left(\tilde{\chi}_{\text{me},n}^{\ell} \ast \tilde{E}^\ell_{\text{av},n}+\eta_0\tilde{\chi}^\ell_{\text{mm},n}\ast\tilde{H}^\ell_{\text{av},n}\right)
	\end{equation}
	and
	\begin{equation}
	\Delta \tilde{H}_n^\ell =\text{j}\frac{k_0}{\eta_0}\left(\tilde{\chi}_{\text{ee},n}^{\ell}\ast\tilde{E}^\ell_{\text{av},n}+\eta_0\tilde{\chi}_{\text{em},n}^{\ell}\ast\tilde{H}^\ell_{\text{av},n}\right),
	\end{equation}
\end{subequations}
where $\Delta$ and the subscript ``$\text{av}$'' refer to the difference and average of the spectral coefficients, respectively (e.g., $\Delta \tilde{E}_n^\ell=\tilde{E}_n^\ell(\rho_\ell)-\tilde{E}_n^{\ell+1}(\rho_\ell)$, and where $\tilde{E}^\ell_{\text{av},n}=[\tilde{E}_n^\ell(\rho_\ell)+\tilde{E}_n^{\ell+1}(\rho_\ell)]/2$ and $\ast$ denotes discrete convolution product with respect to $n$. Equations~\eqref{eq:GSTC_spectral} form a set of $2L(2N+1)$ equations with $2L(2N+1)$ unknown, the expansion coefficients $a_n^\ell$ and $b_n^\ell$ in~\eqref{eq:fields_expansion_appendix}. Note that since the number of regions is $L+1$, the number of expansion modes is $(2N+1)$ and there are $2$ coefficients per region, the number of expansion coefficients is $2(L+1)(2N+1)$, which is greater than the size of the matrix system, $2L(2N+1)$. However, $b_n^1$  and $a_n^{L+1}$ are known quantities: For external (plane-wave) illumination, $b_n^1\neq0$ and $a_n^{L+1}=0$, while for internal (point-source) illumination, $b_n^1=0$ and $a_n^{L+1}\neq0$. Thus, the number of \emph{unknown} coefficients is really $2L(2N+1)$, corresponding to the size of the matrix system.

\subsubsection{External Illumination}\label{sec:outside_illumination}
%
In this case,
\begin{subequations}
	\begin{equation}
	b_n^1=1
	\end{equation}
	and
	\begin{equation}
	a_n^{L+1}=0,
	\end{equation}
\end{subequations}
where $b_n^1\neq 0$ ($\forall n$) corresponds to the expansion of the (assumed) incident plane wave in cylindrical wave functions~\cite{harrington2001time} and $a_n^{L+1}=0$ ($\forall n$) corresponds to the absence of internal illumination.  

Inserting~\eqref{eq:expansion_coefficients} into~\eqref{eq:GSTC_spectral} leads to the matrix equation~\eqref{eq:matrix_equation_outside}, given at the top of next page,
\begin{figure*}[!t]
	\normalsize
	\setcounter{equation}{19}
	\begin{equation}
	\label{eq:matrix_equation_outside}
	\begin{pmatrix}
	&\textbf{P}^{\rho_1}_{\boldsymbol{a}^{1}} &\textbf{P}^{\rho_1}_{\boldsymbol{a}^{2}} &\textbf{P}^{\rho_1}_{\boldsymbol{b}^{2}} &\bf{0} &\bf{0} &\cdots &\bf{0} &\bf{0} &\bf{0}   \\
	&\textbf{Q}^{\rho_1}_{\boldsymbol{a}^{1}} &\textbf{Q}^{\rho_1}_{\boldsymbol{a}^{2}} &\textbf{Q}^{\rho_1}_{\boldsymbol{b}^{2}} &\bf{0} &\bf{0} &\cdots &\bf{0} &\bf{0} &\bf{0}   \\
	&\bf{0} &\textbf{P}^{\rho_2}_{\boldsymbol{a}^{2}} &\textbf{P}^{\rho_2}_{\boldsymbol{b}^{2}} &\textbf{P}^{\rho_2}_{\boldsymbol{a}^{3}} &\textbf{P}^{\rho_2}_{\boldsymbol{b}^{3}} & \cdots &\bf{0} &\bf{0} &\bf{0}   \\
	&\bf{0} &\textbf{Q}^{\rho_2}_{\boldsymbol{a}^{2}} &\textbf{Q}^{\rho_2}_{\boldsymbol{b}^{2}} &\textbf{Q}^{\rho_2}_{\boldsymbol{a}^{3}} &\textbf{Q}^{\rho_2}_{\boldsymbol{b}^{3}} & \cdots &\bf{0} &\bf{0} &\bf{0}   \\
	&\vdots &\vdots &\vdots &\vdots &\vdots & \ddots &\vdots &\vdots &\vdots   \\
	&\bf{0} &\bf{0} &\bf{0} &\bf{0} &\bf{0} &\cdots &\textbf{P}^{\rho_L}_{\boldsymbol{a}^{L}} &\textbf{P}^{\rho_L}_{\boldsymbol{b}^{L}} &\textbf{P}^{\rho_L}_{\boldsymbol{b}^{L+1}} \\
	&\bf{0} &\bf{0} &\bf{0} &\bf{0} &\bf{0} &\cdots &\textbf{Q}^{\rho_L}_{\boldsymbol{a}^{L}} &\textbf{Q}^{\rho_L}_{\boldsymbol{b}^{L}} &\textbf{Q}^{\rho_L}_{\boldsymbol{b}^{L+1}} 
	\end{pmatrix}
	\begin{pmatrix}
	\boldsymbol{a}^1\\
	\boldsymbol{a}^2\\
	\boldsymbol{b}^2\\
	\boldsymbol{a}^3\\
	\boldsymbol{b}^3\\
	\vdots \\
	\boldsymbol{a}^L\\
	\boldsymbol{b}^L\\
	\boldsymbol{b}^{L+1}\\
	\end{pmatrix}
	=
	\begin{pmatrix}
	-\textbf{P}^{\rho_1}_{\boldsymbol{b}^{1}}~\mathds{1} \\
    -\textbf{Q}^{\rho_1}_{\boldsymbol{b}^{1}}~\mathds{1} \\
	\mathds{O} \\
	\mathds{O} \\
	\vdots\\
	\mathds{O}\\
	\mathds{O}\\
	\end{pmatrix},
	\end{equation}
	\hrulefill
	\vspace*{4pt}
	\setcounter{equation}{20}
\end{figure*}
which involves the $(2N+1)\times 1$ vectors
\begin{subequations}\label{eq:coefficients_vectors}
	\begin{equation}
	\boldsymbol{a}^\ell=
	\begin{pmatrix}
	a_{-N}^\ell \\           
	\vdots \\
	a_{N}^\ell
	\end{pmatrix},
	\end{equation}
	\begin{equation}
	\boldsymbol{b}^\ell=
	\begin{pmatrix}
	b_{-N}^\ell \\           
	\vdots \\
	b_{N}^\ell
	\end{pmatrix},
	\end{equation}	
\end{subequations}
containing the unknown expansion coefficients $a_n^\ell$ and $b_n^\ell$, and the $(2N+1)\times(2N+1)$ coefficient matrices
\begin{subequations}\label{eq:coefficients}
	\begin{equation}
	\textbf{P}^{\rho_\ell}_{\boldsymbol{a}^{\ell}} = \left(1-\text{j}\frac{k_0}{2}\tilde{\boldsymbol{\chi}}^\ell_\text{me}\right)\textbf{H}^{\rho_\ell}_{k_\ell} 
	-\frac{k_0\eta_0}{2\eta_\ell}\tilde{\boldsymbol{\chi}}^\ell_\text{mm}\textbf{H}^{\prime \rho_\ell}_{k_\ell},
	\end{equation}
	\begin{equation}
	\textbf{P}^{\rho_\ell}_{\boldsymbol{b}^{\ell}} = \left(1-\text{j}\frac{k_0}{2}\tilde{\boldsymbol{\chi}}^\ell_\text{me}\right)\textbf{J}^{\rho_\ell}_{k_\ell} 	-\frac{k_0\eta_0}{2\eta_\ell}\tilde{\boldsymbol{\chi}}^\ell_\text{mm}\textbf{J}^{\prime\rho_\ell}_{k_\ell},
	\end{equation}
	\begin{equation}
	\textbf{P}^{\rho_\ell}_{\boldsymbol{a}^{\ell+1}} = \left(-1-\text{j}\frac{k_0}{2}\tilde{\boldsymbol{\chi}}^\ell_\text{me}\right)\textbf{H}^{\rho_\ell}_{k_{\ell+1}} 
	-\frac{k_0\eta_0}{2\eta_{\ell+1}}\tilde{\boldsymbol{\chi}}^\ell_\text{mm}\textbf{H}^{\prime \rho_\ell}_{k_{\ell+1}},
	\end{equation}
	\begin{equation}
	\textbf{P}^{\rho_\ell}_{\boldsymbol{b}^{\ell+1}} = \left(-1-\text{j}\frac{k_0}{2}\tilde{\boldsymbol{\chi}}^\ell_\text{me}\right)\textbf{J}^{\rho_\ell}_{k_{\ell+1}}
	-\frac{k_0\eta_0}{2\eta_{\ell+1}}\tilde{\boldsymbol{\chi}}^\ell_\text{mm}\textbf{J}^{\prime\rho_\ell}_{k_{\ell+1}},
	\end{equation}
    \begin{equation}
    \textbf{Q}^{\rho_\ell}_{\boldsymbol{a}^{\ell}} =\frac{\eta_0}{\eta_\ell} \left(1-\text{j}\frac{k_0}{2}\tilde{\boldsymbol{\chi}}^\ell_\text{em}\right)\textbf{H}^{\prime\rho_\ell}_{k_\ell} 
    +\frac{k_0}{2}\tilde{\boldsymbol{\chi}}^\ell_\text{ee}\textbf{H}^{\rho_\ell}_{k_\ell},
    \end{equation}
    \begin{equation}
    \textbf{Q}^{\rho_\ell}_{\boldsymbol{b}^{\ell}} = \frac{\eta_0}{\eta_\ell}\left(1-\text{j}\frac{k_0}{2}\tilde{\boldsymbol{\chi}}^\ell_\text{em}\right)\textbf{J}^{\prime\rho_\ell}_{k_\ell} 	+\frac{k_0}{2}\tilde{\boldsymbol{\chi}}^\ell_\text{ee}\textbf{J}^{\rho_\ell}_{k_\ell},
    \end{equation}
    \begin{equation}
    \textbf{Q}^{\rho_\ell}_{\boldsymbol{a}^{\ell+1}} = -\frac{\eta_0}{\eta_{\ell+1}}\left(1+\text{j}\frac{k_0}{2}\tilde{\boldsymbol{\chi}}^\ell_\text{em}\right)\textbf{H}^{\prime\rho_\ell}_{k_{\ell+1}} 
    +\frac{k_0}{2}\tilde{\boldsymbol{\chi}}^\ell_\text{ee}\textbf{H}^{\rho_\ell}_{k_{\ell+1}}
    \end{equation}
    and
    \begin{equation}
    \textbf{Q}^{\rho_\ell}_{\boldsymbol{b}^{\ell+1}} = -\frac{\eta_0}{\eta_{\ell+1}}\left(1+\text{j}\frac{k_0}{2}\tilde{\boldsymbol{\chi}}^\ell_\text{em}\right)\textbf{J}^{\prime\rho_\ell}_{k_{\ell+1}}
    +\frac{k_0}{2}\tilde{\boldsymbol{\chi}}^\ell_\text{ee}\textbf{J}^{\rho_\ell}_{k_{\ell+1}},
    \end{equation}
\end{subequations}
which involve the $(2N+1)\times(2N+1)$ Toeplitz susceptibility matrix
\begin{subequations}
\begin{equation}
\tilde{\boldsymbol{\chi}}^\ell_{pq}=
\begin{pmatrix}
&\tilde{\chi}_{pq,0}^\ell&\tilde{\chi}_{pq,-1}^\ell &\cdots&\tilde{\chi}_{pq,-2N}^\ell   \\
&\tilde{\chi}_{pq,1}^\ell&\tilde{\chi}_{pq,0}^\ell &\cdots&\tilde{\chi}_{pq,-2N+1}^\ell   \\
&\vdots&\vdots &\ddots&\vdots   \\
&\tilde{\chi}_{pq,2N}^\ell&\tilde{\chi}_{pq,2N-1}^\ell &\cdots&\tilde{\chi}_{pq,0}^\ell  \\ 
\end{pmatrix},
\end{equation}
where $p,q=\text{e,m}$, the $(2N+1)\times(2N+1)$ diagonal Hankel and Bessel functions and their derivative matrices
\begin{equation}
\textbf{H}^{\rho_\ell}_{k_\ell}  = \text{diag}\left(H^{(2)}_n(k_\ell\rho_\ell)\right),
\end{equation}
\begin{equation}
\textbf{H}^{\prime \rho_\ell}_{k_\ell} =\text{diag}\left(H^{(2)\prime}_n(k_\ell\rho_\ell)\right),
\end{equation}
\begin{equation}
\textbf{J}^{\rho_\ell}_{k_\ell}=\text{diag}\left(J_n(k_\ell\rho_\ell)\right)
\end{equation}
and
\begin{equation}
\textbf{J}^{\prime \rho_\ell}_{k_\ell}=\text{diag}\left(J^{\prime}_n(k_\ell\rho_\ell)\right).
\end{equation}
\end{subequations}
the $(2N+1)\times 1$ identity and zero vectors $\mathds{1}$ and $\mathds{O}$, respectively, and the  $(2N+1)\times(2N+1)$ $\bf{0}$ zero matrix. It may be easily verified that the coefficient matrix in~\eqref{eq:matrix_equation_outside} has the dimension $2L(2N+1)\times2L(2N+1)$. Note that the coefficient matrix is a ones diagonal-band matrix because only $a_{\ell}$, $b_{\ell}$, $a_{\ell+1}$ and $b_{\ell+1}$ contribute in the GSTCs~\eqref{eq:GSTC_spectral} for each metasurface $\ell$.   

\subsubsection{Internal Illumination}\label{sec:inside_illumination}
%
In this case,
\begin{subequations}
	\begin{equation}
	b_n^1=0
	\end{equation}
	and
	\begin{equation}
	a_n^{L+1}=-\text{j}^n\frac{k_{L+1}\eta_{L+1}}{4}J_n\left(k_{L+1}\rho^\prime\right)\text{e}^{-\text{j}n\phi^\prime},
	\end{equation}
\end{subequations}
where $b^1_n=0$ ($\forall N$) corresponds to the absence of external illumination and $a_n^{L+1}\neq0$ ($\forall N$) corresponds to the circular-cylindrical wave expansion of the radiated fields of an off-center line source placed at the polar coordinates $(\rho^\prime,\phi^\prime)$~(see~\cite{dudley1994mathematical}.

Eqs.~\eqref{eq:GSTC_spectral} lead to the same matrix system as for the case of external illumination, namely Eq.~\eqref{eq:matrix_equation_outside}, but with the input vector on the right-hand side of the equation replaced by
\begin{equation}
\begin{pmatrix}
\mathds{O} \\
\mathds{O} \\
\mathds{O}\\
\mathds{O}\\
\vdots\\
-\textbf{P}^{\rho_L}_{\boldsymbol{a}^{L+1}}~\boldsymbol{\kappa} \\
-\textbf{Q}^{\rho_L}_{\boldsymbol{a}^{L+1}}~\boldsymbol{\kappa} \\
\end{pmatrix},
\end{equation}
where $\boldsymbol{\kappa}=\boldsymbol{a}^{L+1}$.

\bibliography{NRCloak}

\end{document}